\documentclass{article}
\usepackage[utf8]{inputenc}
\usepackage{amsmath}
\usepackage{mathtools}
\usepackage{amsfonts}
\usepackage{physics}
\usepackage{float}
\usepackage{graphicx}
\usepackage{verbatim}
\usepackage{bbm}
\usepackage{bm}
\usepackage{algorithmic}
\usepackage{algorithm}
\usepackage[round]{natbib}
\usepackage{amsthm} 
\usepackage{algorithmic}
\usepackage{algorithm}
\usepackage{geometry}
\usepackage{mleftright}
\usepackage{subcaption}
\usepackage{tikz}
\usetikzlibrary{calc,shapes,arrows,decorations.pathmorphing,graphs,positioning,backgrounds, chains,arrows.meta}
\usetikzlibrary{decorations.pathreplacing}
\usepackage{multirow}
\DeclareMathOperator{\Var}{Var}

\DeclareMathOperator{\E}{E}

\newcommand{\dU}{U}

\begin{document}



\title{\bfseries \sffamily Bayesian inference for dynamic vine copulas in higher dimensions}

	\date{\small \today}
			\author{Alexander Kreuzer \footnote{Corresponding author: { a.kreuzer@tum.de, Boltzmannstr. 3, 85748 Garching b. München, Germany}}  \and Claudia Czado}
\date{%
	Zentrum Mathematik, Technische Universit\"at M\"unchen\\[2ex]%
	\today
}
			\maketitle
\vspace*{-0.2cm}


\begin{abstract}
We propose a class of dynamic vine copula models. This is an extension of static vine copulas and a generalization of dynamic C-vine and D-vine copulas studied by \cite{almeida2016modeling} and \cite{goel2019analyzing}. Within this class, we allow for time-varying dependence by driving the vine copula parameters with latent AR(1) processes. This modeling approach is very flexible but estimation is not straightforward due to the high-dimensional parameter space. We propose a Bayesian estimation approach, which relies on a novel approximation of the posterior distribution. This approximation allows to use Markov Chain Monte Carlo methods, such as elliptical slice sampling, in a sequential way. In contrast to other Bayesian sequential estimation procedures for vine copula models as proposed by \cite{gruber2015sequential}, there is no need to collapse copula parameters to point estimates before proceeding to the next tree. Thus more information and uncertainty is propagated from lower to higher trees. A simulation study shows satisfactory performance of the Bayesian procedure. This dynamic modeling and inference approach can be applied in various fields, where static vine copulas have already proven to be successful, including environmental sciences, medicine and finance.
Here we study the dependence among 21 exchange rates. For comparison we also estimate a static vine copula model and dynamic C-vine and D-vine  copula models. This comparison shows superior performance of the proposed dynamic vine copula model with respect to one day ahead forecasting accuracy.

\vspace*{0.25cm}
\end{abstract}

\section{Introduction}

Appropriate models for dependence are crucial in many areas. For example, the risk of a portfolio, consisting of several financial assets, is influenced by the dependence among those assets (\cite{embrechts2002correlation}). The multivariate normal distribution is often not sufficient to describe complex dependence structures. It does not allow for tail dependence or asymmetry in the tails.
A more flexible framework is provided by copulas. Sklar's theorem (\cite{sklar1959fonctions}) states that any multivariate distribution can be decomposed into its marginal distributions and the corresponding copula. So the modeling process can be divided into two parts: the marginals and the dependence model. After modeling the margins we can focus entirely on the dependence structure, i.e. on finding an appropriate copula model. In higher dimensions, the class of vine copulas (\cite{bedford2001probability}, \cite{aas2009pair}, \cite{czadoanalyzing}) is especially useful. Vine copulas   are constructed from bivariate copula models. These bivariate copula models can be chosen individually from different classes of copulas, such as elliptical and Archimedean copulas. Due to their flexibility, vine copula models have gained 
huge popularity. They have been successfully applied in many fields, including environmental sciences (\cite{erhardt2015r},  \cite{moller2018vine}), medicine (\cite{barthel2018dependence}) and finance (\cite{brechmann2013risk}, \cite{aas2016pair}, \cite{stubinger2018statistical}).

Although in many of these applications it is assumed that the dependence does not change over time, this assumption is often not appropriate. For example, there is evidence that the correlations between the returns of stocks and bonds change over time (\cite{baele2010determinants}). 
 Popular models for financial data that account for dynamic dependence are multivariate GARCH models with time-varying correlations, such as the DCC-GARCH (\cite{engle2002dynamic}) and multivariate factor stochastic volatility models (\cite{harvey1994multivariate}, \cite{pitt1999time}, \cite{kastner2017efficient}). But these models again rely (conditionally) on  multivariate normal distributions. New models have been proposed to overcome the shortcomings of the multivariate normal distribution and to allow for more flexible time-varying dependence structures. One example for such a model is the dynamic copula model of \cite{oh2018time}.  Another approach to construct dynamic dependence models in higher dimensions is to extend the flexible class of vine copulas.
As already mentioned above, vine copulas are constructed from bivariate copulas. So the vine copula framework allows us to scale dynamic bivariate copula models to arbitrary dimensions. \cite{acar2019flexible} use nonparametric smoothing techniques to allow for time-variation in bivariate copula models and extend this bivariate approach to higher dimensions using vine copulas. \cite{vatter2015generalized} propose a bivariate copula model, where the copula parameters depend on covariates through generalized additive models. Using the vine copula framework, this bivariate model is extended to higher dimensions by \cite{vatter2018generalized}. Similarly the bivariate dynamic copula model as proposed by \cite{almeida2012efficient} and \cite{hafner2012dynamic} is extended to dynamic D-vine copulas by \cite{almeida2016modeling} and later to dynamic C-vine copulas by \cite{goel2019analyzing}.

The bivariate dynamic copula model of \cite{almeida2012efficient} provides a flexible building block by modeling  time-varying dependencies with latent AR(1) processes. However, estimation is no longer straightforward since the likelihood involves high-dimensional integration. 
\cite{goel2019analyzing} follow \cite{almeida2016modeling}, who use a frequentist estimation approach with approximation  of the likelihood utilizing efficient importance sampling (\cite{richard2007efficient}). In this approach, parameters are estimated sequentially  tree by tree. For estimating parameters of higher trees, the parameters of lower trees are fixed at point estimates. Thus uncertainty of parameters in lower trees is ignored and therefore uncertainty quantification cannot be provided.

The paper contains two major contributions:
So far dynamic vine copula models, as generalization of the dynamic bivariate copula model of \cite{almeida2012efficient}, were restricted to D-vine (\cite{almeida2016modeling}) and C-vine (\cite{goel2019analyzing}) structures. 
First we develop an approach  to allow for general regular vine tree structures.
D-vine structures are especially suited to describe temporal dependence. But when it comes to cross-sectional dependence structures, such as the dependence among several stocks, the D-vine structure might be too restrictive. General regular vine tree structures are more flexible and include C-vine and D-vine structures as special cases. 

Second, we present a novel Bayesian estimation approach. To our knowledge Bayesian estimation of vine copula models, including structure selection, was only tackled by \cite{gruber2015sequential} and \cite{gruber2018bayesian}. These approaches only allow for static pair copulas and have not been applied in more than 10 dimensions, while our approach allows for static as well as dynamic pair copulas and can handle higher dimensions.

Our methodology is based on an approximation of the posterior distribution. Approximations to the posterior are also used in variational Bayesian inference (\cite{wainwright2008graphical}) and have become popular since they make estimation feasible in high-dimensional settings. Variational Bayesian approaches assume that the posterior distribution belongs to some family of distributions, such as the multivariate normal distribution. Our approach does not rely on such an assumption. We propose an approximation,  which uses ideas of the frequentist sequential procedure of \cite{dissmann2013selecting} and which allows to estimate pair copulas of one tree independently of each other using Markov Chain Monte Carlo (MCMC) schemes developed in \cite{2019arXiv190210412K}. The MCMC schemes rely on elliptical slice sampling (\cite{murray2010elliptical}) to exploit the underlying autoregressive dependence structure and on an ancillarity-sufficiency interweaving strategy (\cite{yu2011center}).
The posterior approximation also enables the user to run several MCMC chains in parallel leading to faster computation and making the approach applicable to higher dimensions than the approach of \cite{gruber2018bayesian}. Further our Bayesian approach includes pair copula family selection based on a set of candidate families. Here we exploit the fact that for several copula families, there is a one-to-one correspondence between the copula parameter and Kendall's $\tau$. This allows to share the Kendall's $\tau$ parameter among different copula families, which reduces the parameter space and simplifies estimation.
Additionally, our approach also contributes to the selection of sparse models by assessing, whether a pair copula term needs to be modeled dynamically or not. For this the information criteria of \cite{watanabe2010asymptotic} is adapted. Another advantage of our Bayesian approach is that it is not necessary to fix copula parameters at point estimates although our procedure is sequential. Uncertainty of parameter estimates and information in lower trees is no longer ignored, but propagated as we move up to higher trees in the estimation procedure. 
All ideas are investigated through simulation and illustrated with real data.

The outline of the paper is as follows: Section \ref{sec:bblocks} discusses the bivariate building blocks that are needed to construct the dynamic vine copula model. We introduce dynamic and static building blocks and show how to select among them. The selection procedure is illustrated with a small simulation study. 
Section \ref{sec:dynvinecop} introduces the dynamic vine copula model and a novel algorithm for parameter estimation. The performance of the estimation procedure is evaluated with simulated data. In Section \ref{sec:app} we model the dependence among 21 exchange rates. Within this application, we compare the predictive accuracy of the proposed dynamic vine copula model to competitor models: a static vine copula, a dynamic C-vine copula and a dynamic D-vine copula. We conclude with providing ideas for future research in Section \ref{sec:conc}.

\section{Bivariate building blocks for the dynamic vine copula model}
\label{sec:bblocks}

The dynamic vine copula model, we introduce in Section \ref{sec:dynvinecop},  relies on dynamic and static bivariate copula models. These bivariate models are introduced in Sections \ref{sec:bivdynmod} and \ref{sec:bivconstmod}. Section \ref{sec:WAIC} discusses selection among different bivariate copula models.

\subsection{Dynamic bivariate copulas}
\label{sec:bivdynmod}

\subsubsection*{Model specification}
We extend the (time-) dynamic bivariate copula model introduced by \cite{almeida2012efficient} and \cite{hafner2012dynamic} by a model indicator $m$ to allow for Bayesian copula family selection. We consider a set $\mathcal M$ of single-parameter copula families for which there is a one-to-one correspondence between the copula parameter and Kendall's $\tau$. The mapping from the copula parameter  to the corresponding Kendall's $\tau$ is denoted by $g_m(\cdot)$ depending on the copula family $m \in \mathcal M$, e.g. for the Gumbel copula it holds that $g_{Gumbel}(\theta) = 1 -\frac{1}{\theta}$. It is more convenient to work with the corresponding values of Kendall's $\tau$, since parameters of different families may live on different domains and may posses  
different interpretations. Further, we transform Kendall's $\tau$, which is restricted to the interval $(-1,1)$, to the unconstrained scale using the Fisher's Z transformation $F_Z(x)=\frac{1}{2}\log(\frac{1+x}{1-x})$. The transformed Kendall's $\tau$ at time $t$ is denoted by $s_t$ and its dynamics is modeled by an AR(1) process. More precisely, the AR(1) process with states $s_0, \ldots, s_T$, mean $\mu \in \mathbb{R}$, persistence parameter $\phi \in (-1,1)$ and standard deviation parameter $\sigma \in (0, \infty)$ satisfies
\begin{equation}
s_t = \mu + \phi (s_{t-1} - \mu) + \sigma \eta_t
\label{eq:lar}
\end{equation}
for $t = 1, \ldots, T$. Here we assume that the errors are $\eta_t \sim N(0,1) \text{ iid}$ and the initial distribution is $s_0|\mu, \phi, \sigma \sim N\left(\mu, \frac{\sigma^2}{1-\phi^2}\right)$. The state $s_t$ is mapped to the parameter $\theta^m_t$ of the copula family $m$ at time $t$ as follows
\begin{equation}
\theta_t^m = g^{-1}_m(F_Z^{-1}(s_t)).
\label{eq:dynpar}
\end{equation}
We now study the following model for $T$  bivariate random vectors $(U_{t1}, U_{t2})_{t=1, \ldots, T} \in [0,1]^{T\times 2}$
\begin{equation}
(U_{t1}, U_{t2})|m,s_t \sim c^m(u_{t1}, u_{t2}; \theta_t^m) \text{ independently}, 
\label{eq:bivdyn}
\end{equation}
for $t = 1, \ldots, T$. Here $c^m(\cdot, \cdot;\theta_t^m)$ is the bivariate density of copula family $m$ with parameter $\theta_t^m$ as specified in \eqref{eq:lar} and \eqref{eq:dynpar}. The copula parameter $\theta_t^m$ is a function of the model indicator $m$ and the state $s_t$. The state $s_t$ has the same interpretation for different copula families as the Fisher's Z transform of the corresponding Kendall's $\tau$ value. This allows us to share the parameters $\boldsymbol {s_{0:T}} = (s_0, \ldots, s_T), \mu, \phi, \sigma$ among different copula families, which keeps the parameter space smaller and simplifies estimation. More details about parameter sharing are given in Appendix \ref{sec:app_ps}.

A Bayesian model specification is complete by introducing priors for the model parameters.
Let $\varphi(\cdot|\mu_{Normal},\sigma_{Normal}^2)$ denote the density of the univariate normal distribution with mean $\mu_{Normal}$ and variance $\sigma_{Normal}^2$.
This allows us to express the prior for $\boldsymbol {s_{0:T}}$ conditional on $\mu, \phi$ and $\sigma$, implied by the AR(1) process in \eqref{eq:lar}, as
\begin{equation}
\pi(\boldsymbol {s_{0:T}}|\mu, \phi, \sigma) = \varphi(s_0|\mu, {\sigma^2}({{1-\phi^2}})^{-1}) \prod_{t=1}^T \varphi(s_t|\mu + \phi(s_{t-1} - \mu), \sigma^2).
\label{eq:priorAR}
\end{equation}
Further, we assume that
\begin{equation}
\mu \sim N(0,100), ~\frac{\phi+1}{2} \sim Beta(5,1.5), ~\sigma^2 \sim Gamma\left(\frac{1}{2},\frac{1}{2}\right)
\label{eq:prior}
\end{equation}
as in \cite{2019arXiv190210412K}. These are the same priors that \cite{kastner2016dealing} recommend for the latent AR(1)  process of the stochastic volatility model. For $\mu$ we utilize a vague prior. But for $\phi$ we use a rather informative prior, which gives more prior probability to larger values within $(-1,1)$. Thus we favor trajectories of Kendall's $\tau$, where there is positive dependence among two subsequent time points ($\text{cor}(s_t, s_{t-1})=\phi$). The Gamma prior for the variance parameter $\sigma^2$ is different to the frequently used inverse Gamma prior. The Gamma prior has more mass close to zero compared to the inverse Gamma prior. So we give more probability to smoother trajectories that do not oscillate a lot.
For $m \in \mathcal{M}$ we assume a discrete uniform prior, i.e.
\begin{equation}
\pi(m) = \frac{1}{|\mathcal{M}|}.
\label{eq:priorI}
\end{equation}
We further assume prior independence among the parameters $\mu, \phi, \sigma$ and $m$.

\subsubsection*{Bayesian inference}

Model selection procedures often have to deal with model specific parameters. They might have different dimensions. A popular approach in this context is reversible jump MCMC (see \cite{green1995reversible}), which requires dimension matching. \cite{min2011bayesian} and \cite{gruber2015sequential} use reversible jump MCMC for selection among vine copula models, while \cite{tan2019bayesian} apply it for model selection among static single factor copula models. The way we constructed our model, dimension matching is not needed, since we share the parameters $s_0, \ldots , s_T, \mu$, $\phi$ and $\sigma$ among different models. This allows us to use an efficient Gibbs approach as outlined in the following.

 Here the likelihood given the data $U = (u_{t1},u_{t2})_{t=1, \ldots, T}$ can be expressed as
\begin{equation}
\ell(\mu, \phi, \sigma, \boldsymbol {s_{0:T}},m|U) =\prod_{t=1}^T \ell_t(s_t,m|u_{t1},u_{t2})=\prod_{t=1}^T c^m(u_{t1},u_{t2};\theta_t^m) .
\label{eq:lik}
\end{equation}
The quantity $\ell_t(s_t,m|u_{t1},u_{t2}) = c^m(u_{t1},u_{t2};\theta_t^m)$ is the contribution to the likelihood at time $t$ and $\theta_t^m = g_m^{-1}(F_Z^{-1}(s_t))$.

We now employ a Gibbs sampler for parameter estimation. The indicator $m$ is sampled from its full conditional, given by
\begin{equation}
\begin{split}
P(m|U,\mu, \phi, \sigma, \boldsymbol {s_{0:T}})
  &=\frac{f(U|\mu, \phi, \sigma, \boldsymbol {s_{0:T}}, m)  \pi(\mu, \phi, \sigma, \boldsymbol {s_{0:T}}) \pi(m)}{\sum_{ m' \in \mathcal{M}}f(U|\mu, \phi, \sigma, \boldsymbol {s_{0:T}},m')\pi(\mu, \phi, \sigma, \boldsymbol {s_{0:T}}) \pi( m')}\\
&= \frac{\ell(\mu, \phi, \sigma, \boldsymbol {s_{0:T}}, m|U) }{\sum_{m' \in \mathcal{M}}\ell(\mu, \phi, \sigma, \boldsymbol {s_{0:T}},m'|U)} \\
&= \frac{\prod_{t=1}^T 
   c^{m}(u_{t1},u_{t2};\theta_{t}^m)}{\sum_{ m' \in \mathcal{M}} \prod_{t=1}^T 
  c^{m'}(u_{t1},u_{t2};\theta_{t}^{m'})},
\end{split}
\label{eq:fcond_I}
\end{equation}
where $\pi(\mu, \phi, \sigma, \boldsymbol {s_{0:T}}) = \pi(\boldsymbol {s_{0:T}}| \mu, \phi, \sigma) \pi(\mu) \pi(\phi) \pi(\sigma)$  and $\pi(\cdot)$ as specified in \eqref{eq:priorAR},  \eqref{eq:prior} and  \eqref{eq:priorI}.
Here the presence of the shared parameters simplifies the updates of the copula family indicator $m$ considerably. To sample $\mu, \phi, \sigma, \boldsymbol {s_{0:T}}$ conditioned on the indicator $m$, we use the same approach as described in \cite{2019arXiv190210412K} for multivariate state space models with a univariate autoregressive state equation. This sampler relies on an interweaving strategy (\cite{yu2011center}), elliptical slice sampling (\cite{murray2010elliptical}) and adaptive Metropolis-Hastings updates (\cite{garthwaite2016adaptive}).

\subsection{Static bivariate copulas}
\label{sec:bivconstmod}
\subsubsection*{Model specification}

To allow for static (time-constant) copulas we consider a static state $s \in \mathbb{R}$ which is mapped to the copula parameter similar to \eqref{eq:dynpar}, i.e. 
\begin{equation}
\theta^m = g^{-1}_m(F_Z^{-1}(s))
\label{eq:cpar}
\end{equation}
for a copula family $m \in  \mathcal{M}$. We assume that $T$ bivariate random vectors $(U_{t1},U_{t2})_{t=1, \ldots, T}$ are generated as follows 
\begin{equation}
\begin{split}
&(U_{t1}, U_{t2})|m,s \sim c^m(\cdot, \cdot; \theta^m), \text{ independently}. \\ 
\end{split}
\label{eq:ccop}
\end{equation}
The prior for the state parameter $s$ is chosen such that the corresponding Kendall's $\tau$ is uniformly distributed on $(-1,1)$. The prior for $m$ is chosen as above in \eqref{eq:priorI}. The priors reflect the fact that we do not have any prior information about the parameters. 
\subsubsection*{Bayesian inference}

The parameters of this reduced model are also estimated utilizing a Gibbs sampling approach. Here we sample $m|U,s$ directly from its full conditional, which can be derived similar to \eqref{eq:fcond_I}. The parameter $s$ is updated conditional on $(U,m)$ with random walk Metropolis-Hasting with Gaussian proposal and adaptive proposal variance as in \cite{garthwaite2016adaptive}.

\subsection{Model selection among dynamic and static pair copulas using the widely applicable information criteria (WAIC)}
\label{sec:WAIC}
One might be interested in how the dynamic copula model compares to the static copula model and to the independence model. 
We refer to these model classes as dynamic, static and zero dependence, respectively.
To also allow for bivariate static and independence copulas is especially important, if the bivariate dynamic copula is used as a building block for vine copula models in higher dimensions. For vine copula models, independence copulas might be useful in higher trees to avoid overfitting. 

Bayes factors or the commonly used information critera AIC and BIC are here not tractable choices for selecting the type of dependence (dynamic, static or zero), since their evaluation would require high-dimensional integration. Instead, we rely on the widely applicable information criteria (WAIC) introduced by \cite{watanabe2010asymptotic}. For the proposed dynamic copula model it is given by

\begin{equation}
\text{WAIC}=-2 \left(  \sum_{t=1}^T\ln(\E(\ell_t(s_t,m|u_{t1},u_{t2}))) -  \sum_{t=1}^T \Var \left(\ln(\ell_t(s_t,m|u_{t1},u_{t2}))\right) \right)
\end{equation}
with $\ell_t$ as in \eqref{eq:lik}.  The expectation and variance are taken with respect to $P(s_t, m)$, the probability measure of the posterior distribution of $s_t$ and $m$, i.e. $$\E(\ell_t(s_t,m|u_{t1},u_{t2})) = \int_{\mathbb{R} \times \mathcal{M}} \ell_t(s_t,m|u_{t1},u_{t2}) dP(s_t,m)$$ and $$\Var(\ln(\ell_t(s_t,m|u_{t1},u_{t2}))) = \int_{\mathbb{R} \times \mathcal{M}} (\ln(\ell_t(s_t,m|u_{t1},u_{t2})))^2 dP(s_t,m) - \left(\E(\ln(\ell_t(s_t,m|u_{t1},u_{t2})))\right)^2.$$
 The WAIC can be seen as a Bayesian version of the AIC, where $\sum_{t=1}^T\Var(\ln(\ell_t(s_t,m|u_{t1},u_{t2}))$ is used as a penalty instead of the number of parameters.

For $R$ observed quantities $(x^r)_{r=1, \ldots, R}$ we denote by $ \hat{E}((x^r)_{r=1, \ldots, R})=\frac{1}{R}\sum_{r=1}^Rx^r$ the sample mean and by $\widehat{\Var}((x^r)_{r=1, \ldots, R})=\frac{1}{R-1}\sum_{r=1}^R(x^r -\hat{E}((x^r)_{r=1, \ldots, R}))^2$ the sample variance.
Following \cite{vehtari2017practical}, the $\text{WAIC}$ can be estimated from $R$ samples of the posterior distribution $ (s_t^1, m^1), \ldots, (s_t^R, m^R)$ with $t=1, \ldots, T$, by

\begin{equation}
\widehat{\text{WAIC}}= -2\left(\sum_{t=1}^T \ln( \hat E((\ell_t^r)_{r=1, \ldots, R})    ) -\sum_{t=1}^T \widehat{\Var} \Big(  (\ln(\ell_t^r))_{r=1, \ldots, R}\Big)        \right), 
\end{equation}
where $\ell_t^r = \ell_t(s_t^r, m^r|u_{t1},u_{t2})$.
By setting $$\widehat{\text{WAIC}_t} = -2 \left( \ln( \hat E((\ell_t^r)_{r=1, \ldots, R})    )  - \widehat{\Var} \Big(  (\ln(\ell_t^r))_{r=1, \ldots, R}\Big) \right),  $$ we can express $\widehat{\text{WAIC}}$ as $\widehat{\text{WAIC}}= \sum_{t=1}^T \widehat{\text{WAIC}_t}$. 

To compare between two models with estimated WAIC values $\widehat{\text{WAIC}}^A$ and $\widehat{\text{WAIC}}^B$,  \cite{vehtari2017practical} suggest to consider the difference in the estimated WAIC given by
\begin{equation}
\widehat{\text{WAIC}}^A - \widehat{\text{WAIC}}^B = \sum_{t=1}^T (\widehat{\text{WAIC}_t}^A - \widehat{\text{WAIC}_t}^B)
\end{equation}
with corresponding standard error estimate
\begin{equation}
\hat{se}(\widehat{\text{WAIC}}^A - \widehat{\text{WAIC}}^B)= \sqrt{T \cdot \widehat{\text{VAR}}((\widehat{\text{WAIC}_t}^A - \widehat{\text{WAIC}_t}^B)_{t=1, \ldots, T})}.
\label{eq:se_waic}
\end{equation}
To estimate the standard error, \cite{vehtari2017practical} assume independence among the components $\widehat{\text{WAIC}}_t, t=1, \ldots, T$.  For the static copula model we use $\ell_t(s,m|u_{t1},u_{t2}) = c^m(u_{t1},u_{t2};\theta^m)$. Further, WAIC is zero for the independence copula.
In our framework the dynamic model is considered to be more complex than the static model. Similarly, the static and the dynamic model are considered to be more complex than the independence model. Here, we select the more complex model if its estimated WAIC is at least 2 standard errors smaller than the WAIC of the other model.

\tikzstyle{ClassicalNode} = [rectangle, fill = lightgray!43, draw = black, text = black, align = center]
\tikzstyle{TreeLabels} = [draw = none, fill = none, text = black, font = \bf]
\tikzstyle{DummyNode}  = [draw = none, fill = none, text = white]
\tikzset{
	block/.style = {rectangle, draw,  text width=9.0cm,
		line width=.5pt,minimum height=1.75cm},
	block_half/.style = {block, text width=5cm},
}
\newcommand{\yshift}{-1cm}
\newcommand{\yshiftlabel}{-0.1cm}
\newcommand{\labelsize}{\small}
\begin{figure}[h!]
	\centering
	\begin{tikzpicture}	[every node/.style = ClassicalNode, node distance = .5cm, font = \normalsize, minimum height=1.5cm, minimum width=9cm]
	
	\node[text width=4cm] (Data) {
		\textbf{Data:} $U \in [0,1]^{T\times 2}$ 
	};
	\node [below = of Data, xshift = -6cm, minimum width=0cm] (dm) {
	\textbf{Dynamic model:} \\
	\textbf{(dynamic dependence)}\\
	Estimate a bivariate dynamic  \\
	copula model as explained  \\
	in Section \ref{sec:bivdynmod}. This procedure \\
	includes Bayesian copula family \\
	 selection.};
	 
\node [below = of Data, xshift = -0.5cm, minimum width=0cm] (cm) {
	\textbf{Static model:}\\
	\textbf{(static dependence)}\\
	Estimate a bivariate  \\
	 static copula model as \\
	 explained in Section \ref{sec:bivconstmod} .\\
	 This procedure includes\\
	 Bayesian copula family \\
	  selection.};	 
	 
\node [below = of Data, xshift = 4cm, minimum width=0cm] (im) {
	\textbf{Independence model:}\\
	\textbf{(zero dependence)}\\
	Here is no estimation \\
	 required.\\ 
	 \\
	 \\
	 };	 
	\node [below = of cm, xshift = 0.5cm, minimum width=0cm] (waic) {
	\textbf{Select the type of dependence:}\\
	Use WAIC as explained above to select among the dynamic,\\
	 the static and the zero dependence.} ;	
	\draw[-{Latex[scale=1.25]}] (Data) to node[draw=none, fill = none] {} (dm);
		\draw[-{Latex[scale=1.25]}] (Data) to node[draw=none, fill = none] {} (im);
	\draw[-{Latex[scale=1.25]}] (Data) to node[draw=none, fill = none] {} (cm);

	\draw[-{Latex[scale=1.25]}] (dm) to node[draw=none, fill = none] {} (waic);
		\draw[-{Latex[scale=1.25]}] (im) to node[draw=none, fill = none] {} (waic);
	\draw[-{Latex[scale=1.25]}] (cm) to node[draw=none, fill = none] {} (waic);
	\end{tikzpicture}
	\caption{Model selection procedure for bivariate copula models.}
	\label{fig:flowchart_sel}
\end{figure}
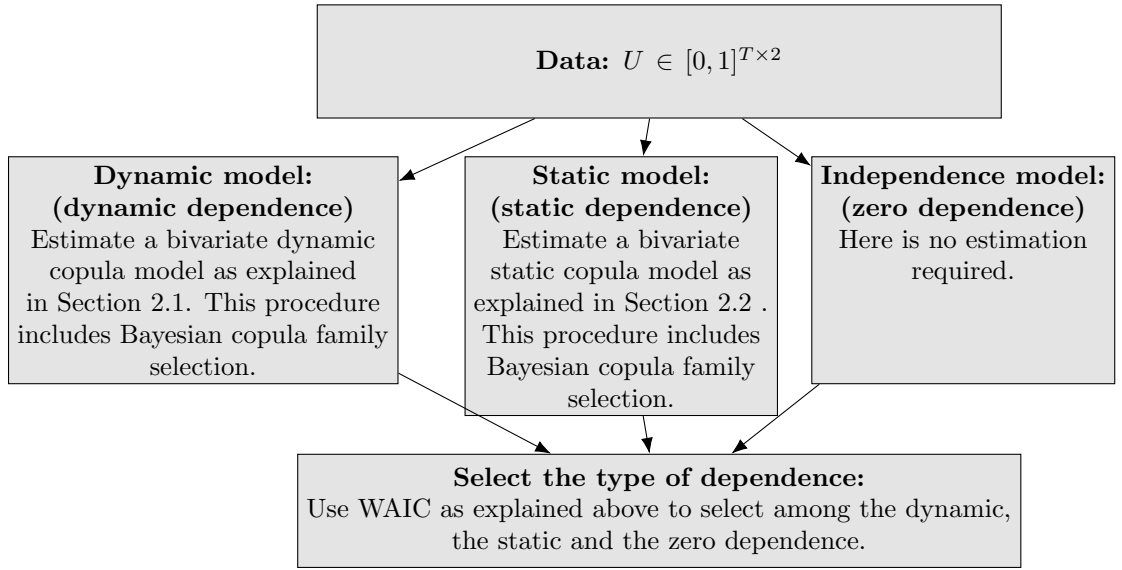

Alternatively we could have incorporated the independence, the static and the dynamic copula within one sampler. In this case we would need to move between models with different dimensions by employing e.g. reversible jump MCMC. But proposing moves efficiently from the parameter free independence copula or from the static copula to dynamic copulas with more than $T$ parameters is difficult and chains might take very long to converge.   
Another alternative is to select among all models, including family choices, with ${\text{WAIC}}$. But this would require to estimate a dynamic bivariate copula model for each copula family, which is computationally expensive.
So we propose to use the Gibbs sampler to move between models with the same dimension, where the parameters can be shared among the different models. The ${\text{WAIC}}$ is used to select between models with different parameter dimensions where parameters can not be shared. The whole selection procedure for the bivariate copula models is visualized in Figure \ref{fig:flowchart_sel}.

\subsection{Simulation study}
We conduct a small simulation study to investigate the ability of WAIC to select the type of dependence. We consider five scenarios specified in Table \ref{tab:sim1_paraspec}. In Scenarios 1 and 2 we simulate from the dynamic model specified in \eqref{eq:bivdyn}, in Scenarios 3 and 4 from the static one specified in \eqref{eq:ccop} and in Scenario 5 from the bivariate independence copula. For each scenario we simulate $T=1000$ observations. Based on these observations we fit the dynamic and the static model and select among different models with WAIC as explained in Section \ref{sec:WAIC}. We consider the following set for copula family selection $\mathcal{M}=\{${Independence, Gaussian, Student t(df=4), eClayton, eGumbel}$\}$. Here eClayton and eGumbel are extended Clayton and Gumbel copulas that also allow for negative Kendall's $\tau$ values as in \cite{2019arXiv190210412K}. We repeat the simulations for each scenario 100 times. From Table \ref{tab:sim1_dyn} we see that in each scenario, the correct type of dependence was selected at least in 84 out of 100 cases. The correct family was selected in at least 98 out of 100 cases according to Table \ref{tab:sim1_dyn}. We conclude that our Bayesian family selection procedure performs well and that WAIC can be utilized to select the appropriate type of dependence.

\begin{table}[H]
\centering
\begin{tabular}{l|ll|ll|l}
  \hline
 & \multicolumn{2}{c} {Dynamic} & \multicolumn{2}{c}{Static} & Independence \\
 & Scenario 1 &Scenario 2 &Scenario 3 &Scenario 4 &Scenario 5 \\   \hline
 $m$ & Gaussian & eClayton &  Student t(df=4) & eGumbel  & Independence \\ 
$\mu$ & 0.4 & 0.4 &  &  &  \\ 
$\phi$   & 0.95 & 0.8 &  &  &  \\ 
$\sigma$   & 0.1 & 0.2 &  &  &  \\ 
$s$ &  & &  1 & 0.4  &  \\ 
   \hline
\end{tabular}
\caption{Parameter specification for the simulation study for bivariate copula models.}
\label{tab:sim1_paraspec}
\end{table}

\begin{table}[ht]
\centering
\begin{tabular}{ll|rr|rr|r}
  \hline
  &   & \multicolumn{2}{c} {Dynamic} & \multicolumn{2}{c}{Static} & Independence \\

 & Scenario &  1 & 2 & 3 & 4 & 5 \\ 
  \hline
Copula family &Independence &   0 &   0 &   0 &   0 &  \textbf{100} \\ 
&Gaussian &  \textbf{100} &   2 &   1 &   0 &   0 \\ 
&Student t(df=4) &  0 &   0 &  \textbf{99} &   0 &   0 \\ 
&eClayton &   0 & \textbf{98} &   0 &   0 &   0 \\ 
&eGumbel &   0 &   0 &   0 & \textbf{100} &   0 \\ 
   \hline

  \hline
Type of dependence &Dynamic & \textbf{100} & \textbf{100} &  16 &   0 &   1 \\ 
&Static &   0 &   0 &  \textbf{84} & \textbf{100} &   0 \\ 
&Zero &   0 &   0 &   0 &   0 &  \textbf{100} \\ 
   \hline
\end{tabular}
\caption{For each scenario of the simulation study, we show how often the different copula families were selected and how often each type of dependence (dynamic, static and zero) was selected.
The selected copula family is the marginal posterior mode estimate of $m$, i.e. the family that occurs most frequently among the posterior samples for $m$. The true copula family and the true type of dependence, which we used for simulation, is marked in bold.
}
\label{tab:sim1_dyn}
\end{table}

\section{Dynamic vine copulas}
\label{sec:dynvinecop}
\subsection{Model specification}
\label{sec:vinemodelspec}

Vine copulas (\cite{bedford2001probability}, \cite{aas2009pair}, \cite{joe2014dependence}, \cite{czadoanalyzing}) are a popular class in dependence modeling. They allow for great flexibility by constructing a density of arbitrary dimension from two-dimensional densities. 

For this construction, vine copulas are represented as graphical models.
A regular vine (R-vine) tree sequence is a sequence of trees $\mathcal{V} = T_1, \ldots, T_{d-1}$ satisfying the following conditions
\begin{itemize}
\item each tree is connected,
\item $T_1$ is a tree with nodes $N_1 = \{1, \ldots d\}$ and set of edges $E_1$,
\item  $T_j$ with $j \geq 2$ is a tree with nodes $N_j = E_{j-1}$ and edges $E_j$ and
\item for $j = 2, \ldots d-1$ and $\{a,b\} \in E_j$, it holds that $a$ and $b$ as edges in tree $T_{j-1}$ share a common node (proximity condition).
\end{itemize}
Furthermore, the complete union of an edge $e \in E_i$ is defined  as
\begin{equation*}
\boldsymbol {A_e} \coloneqq \{j \in N_1| \exists e_1 \in E_1, \ldots, e_{i-1} \in E_{i-1}; j \in e_1 \in \ldots \in e_{i-1} \in e       \}.
\end{equation*}
The conditioning set of edge $e = \{a,b\}$ is obtained as
\begin{equation*}
\boldsymbol {D_e} \coloneqq \boldsymbol {A_a} \cap \boldsymbol {A_b} ,
\end{equation*}
and the conditioned sets are given by
\begin{equation*}
a_e \coloneqq \boldsymbol {A_a} \setminus \boldsymbol {D_e}, \hspace*{0.2cm} b_e \coloneqq \boldsymbol {A_b} \setminus \boldsymbol {D_e} .
\end{equation*}
Bivariate copulas of conditional distributions of the marginally uniformly distributed random vector $(U_1, \ldots, U_d)$ can be identified with the conditioning and the conditioned sets. For an edge $e$ we denote by $c_{a_e,b_e;\boldsymbol {D_e}}$ the density of the bivariate copula of $(U_{a_e} U_{b_e})|\boldsymbol{U_{\boldsymbol {D_e}}} =  \boldsymbol{u_{\boldsymbol {D_e}}}$, where $\boldsymbol{u_{F}} = (u_i)_{i \in F}$ for a set $F$. Many researchers assume that $c_{a_e,b_e;\boldsymbol {D_e}}$ does not depend on $\boldsymbol{u_{\boldsymbol {D_e}}}$, which is called the simplifying assumption (\cite{haff2010simplified}, \cite{stoeber2013simplified}). This assumption allows for sequential estimation and selection of vine copula models (\cite{brechmann2013risk}, \cite{dissmann2013selecting}).

Based on these graphical definitions, \cite{bedford2001probability} build a $d$-dimensional vine copula model with joint density
\begin{equation}
c(u_1, \ldots u_d) =\prod_{i=1}^{d-1} \prod_{e \in E_i} c_{a_e,b_e;\boldsymbol {D_e}}(u_{a_e|\boldsymbol {D_e}}, u_{b_e|\boldsymbol {D_e}}).
\label{eq:pcc}
\end{equation}
It is a simplified vine copula model, since $c_{a_e, b_e;\boldsymbol {D_e}}$ does not depend on the conditioning value $\boldsymbol{u_{D_e}}$.
Here, $u_{a_e|\boldsymbol {D_e}}$ and $u_{b_e|\boldsymbol {D_e}}$ are called pseudo data. They are obtained as $u_{a_e|\boldsymbol {D_e}}=$ \\ $C_{a_e|\boldsymbol {D_e}}(u_{a_e}| \boldsymbol{u_{D_e}})$ and $u_{b_e|\boldsymbol {D_e}}=$ $C_{b_e|\boldsymbol {D_e}}(u_{b_e}| \boldsymbol{u_{D_e}})$, where $C_{a_e|\boldsymbol {D_e}}$ and $C_{b_e|\boldsymbol {D_e}}$ are the conditional distribution functions of $U_{a_e}| \boldsymbol{U_{D_e}} =  \boldsymbol{u_{D_e}} $ and  $ U_{b_e}| \boldsymbol{U_{D_e}} =  \boldsymbol{u_{D_e}}$, respectively. In the first tree $\boldsymbol {D_e}$ is the empty set and the pseudo data of the first tree is just $u_1, \ldots u_d$.  Note that the class of simplified vine copulas is broad, including multivariate Gaussian and Student t copulas (\cite{joe2014dependence}, Chapter 3).

To evaluate the conditional distribution functions and to obtain the corresponding pseudo data for all trees, the ${h}$ functions (\cite{aas2009pair}) for an edge $e$ are defined as 
\begin{equation}
\begin{split}
{h}_{a_e|b_e;\boldsymbol {D_e}}(u_{1}|u_{2}) &= \frac{d}{du_2} C_{a_e,b_e;\boldsymbol {D_e}}(u_1,u_2)\\
{h}_{b_e|a_e;\boldsymbol {D_e}}(u_{2}|u_{1}) &= \frac{d}{du_1} C_{a_e,b_e;\boldsymbol {D_e}}(u_1,u_2).
\end{split}
\label{eq:hdef}
\end{equation}
If the copula $C_{a_e,b_e;\boldsymbol {D_e}}$ depends on a set of parameters $\boldsymbol \delta$ we write
$
{h}_{a_e|b_e;\boldsymbol {D_e}}(u_{1}|u_{2};\boldsymbol \delta)$
and
$
{h}_{b_e|a_e;\boldsymbol {D_e}}(u_{2}|u_{1};\boldsymbol \delta) .
$
Based on pseudo data $u_{a_e|\boldsymbol {D_e}}$  and $u_{b_e|\boldsymbol {D_e}}$, we obtain pseudo data for the next tree as
\begin{equation}
\begin{split}
u_{a_e|b_e \cup \boldsymbol {D_e}} = {h}_{a_e|b_e;\boldsymbol {D_e}}(u_{a_e|\boldsymbol {D_e}}|u_{b_e|\boldsymbol {D_e}}), \\
u_{b_e|a_e \cup \boldsymbol {D_e}} = {h}_{b_e|a_e;\boldsymbol {D_e}}(u_{b_e|\boldsymbol {D_e}}|u_{a_e|\boldsymbol {D_e}}).
\end{split}
\label{eq:hfunc}
\end{equation}
So for the calculation of pseudo data only bivariate copulas from lower trees are involved.

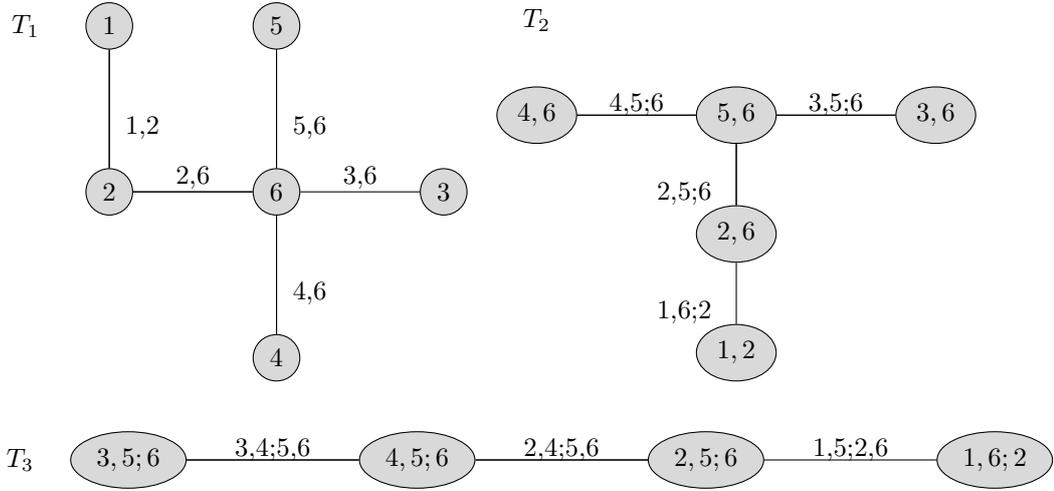
\begin{figure}[H]

\begin{minipage}[t]{.45\textwidth}
     \centering
\begin{tikzpicture}[scale=2.2]
\tikzstyle{every node}=[draw,shape=circle,fill=gray!30];
\tikzstyle{TreeLabels} = [draw = none, fill = none, text = black]
\newcommand{\xshiftNodes}{0.5*\linewidth}
\newcommand{\yshiftLabels}{-.25cm}  
\node (v5) at (1,2) {$5$};
\node (v1) at (0,2) {$1$};
\node (v2) at (0,1) {$2$};
\node (v6) at (1,1) {$6$};
\node (v3) at (2,1) {$3$};
\node (v4) at (1,0) {$4$};
\node[TreeLabels] (v7) at (-0.5,2) {$T_1$};
\draw (v1) -- (v2)
(v2) -- (v6)
(v6) -- (v5)
(v4) -- (v6);
(v6) -- (v3);
\draw (v1) to node[draw=none, fill = none, right, yshift = \yshiftLabels] {1,2} (v2);
\draw (v2) to node[draw=none, fill = none, above, yshift = \yshiftLabels] {2,6} (v6);
\draw (v5) to node[draw=none, fill = none, right, yshift = \yshiftLabels] {5,6} (v6);
\draw (v4) to node[draw=none, fill = none, right, yshift = \yshiftLabels] {4,6} (v6);
\draw (v3) to node[draw=none, fill = none, above, yshift = \yshiftLabels] {3,6} (v6);
\end{tikzpicture}
\vspace*{.5cm}
\end{minipage}
\begin{minipage}[t]{.45\textwidth}
     \centering
\begin{tikzpicture}[scale=2.1]
\tikzstyle{every node}=[draw,shape=ellipse,fill=gray!30];
\tikzstyle{TreeLabels} = [draw = none, fill = none, text = black]
\newcommand{\xshiftNodes}{0.5*\linewidth}
\newcommand{\yshiftLabels}{-.25cm}  
\node (v46) at (0,1) {$4,6$};
\node (v56) at (1.25,1) {$5,6$};
\node (v36) at (2.5,1) {$3,6$};
\node (v26) at (1.25,0.25) {$2,6$};
\node (v12) at (1.25,-0.5) {$1,2$};
\node[TreeLabels] (v7) at (0,1.6) {$T_2$};
\draw (v46) -- (v56)
(v56) -- (v36)
(v26) -- (v56);
\draw (v46) to node[draw=none, fill = none, above, yshift = \yshiftLabels] {4,5;6} (v56);
\draw (v36) to node[draw=none, fill = none, above, yshift = \yshiftLabels] {3,5;6} (v56);
\draw (v26) to node[draw=none, fill = none, left, yshift = \yshiftLabels] {2,5;6} (v56);
\draw (v12) to node[draw=none, fill = none, left, yshift = \yshiftLabels] {1,6;2} (v26);
\end{tikzpicture}
\vspace*{.5cm}
\end{minipage}

\begin{tikzpicture}[scale=1.9]
\tikzstyle{every node}=[draw,shape=ellipse,fill=gray!30];
\tikzstyle{TreeLabels} = [draw = none, fill = none, text = black]
\newcommand{\xshiftNodes}{0.5*\linewidth}
\newcommand{\yshiftLabels}{-.25cm}  
\node (v356) at (0,0) {$3,5;6$};
\node (v456) at (2,0) {$4,5;6$};
\node (v256) at (4,0) {$2,5;6$};
\node (v162) at (6,0) {$1,6;2$};
\node[TreeLabels] (v4) at (-0.75,0) {$T_3$};
\draw (v356) -- (v456)
(v456) -- (v256);
(v256) -- (v162);
\draw (v356) to node[draw=none, fill = none, above, yshift = \yshiftLabels] {3,4;5,6} (v456);
\draw (v456) to node[draw=none, fill = none, above, yshift = \yshiftLabels] {2,4;5,6} (v256);
\draw (v256) to node[draw=none, fill = none, above, yshift = \yshiftLabels] {1,5;2,6} (v162);
\end{tikzpicture}
\vspace*{.5cm}
\caption{Tree structure of a vine copula model}
\label{fig:treestruct}
\end{figure}

Figure \ref{fig:treestruct} visualizes the first three trees of a six-dimensional vine copula. Assuming that there are only independence copulas in trees higher than Tree 3, the associated density is given by
\begin{equation}
\begin{split}
c(u_1, \ldots, u_6) =&  c_{12}(u_1, u_2) \cdot c_{26}(u_2,u_6) \cdot c_{36}(u_3,u_6) \cdot c_{46}(u_4,u_6) \cdot c_{56}(u_5,u_6) \\
\cdot  & c_{45;6}(u_{4|6}, u_{5|6}) \cdot c_{35;6}(u_{3|6}, u_{5|6}) \cdot c_{25;6}(u_{2|6}, u_{5|6}) \cdot c_{16;2}(u_{1|2}, u_{6|2}) \\
\cdot & c_{3,4;5,6}(u_{3|5,6},u_{4|5,6}) \cdot c_{2,4;5,6}(u_{2|5,6},u_{4|5,6}) \cdot c_{1,5;2,6}(u_{1|2,6},u_{5|2,6}).
\end{split}
\end{equation}
The pseudo data of the vine copula model can be determined as outlined in \eqref{eq:hfunc}. E.g.  $u_{4|6} = h_{4|6}(u_4|u_6), u_{5|6}=h_{5|6}(u_5|u_6)$ and $u_{4|56}=h_{4|5;6}(u_{4|6}|u_{5|6})$ $=h_{4|5;6}(h_{4|6}(u_4|u_6)|h_{5|6}(u_5|u_6))$. 

Such vine copulas, where pair copulas above a certain tree level are set to the independence copula, are called truncated (\cite{brechmann2012truncated}). With truncation we can achieve different levels of sparsity.

Within the vine copula framework, we only need to specify bivariate copulas and can scale to arbitrary dimensions. Here, we replace each bivariate copula in \eqref{eq:pcc} by the dynamic bivariate copula model specified in \eqref{eq:bivdyn}, the static bivariate copula model specified in \eqref{eq:ccop} or the independence copula. We denote by $E^{dyn}_i, E^{static}_i$ and $E^{ind}_i$ the set of edges in the $i$-th tree where the corresponding pair copulas are dynamic, static or independence copulas, respectively. For each edge $e \in E^{dyn}_i, i=1, \ldots, d-1$ we have a corresponding family indicator $m_e$ and a corresponding latent AR(1) process given by 
\begin{equation}
s_{t,e}=\mu_{e} + \phi_{e} (s_{t-1,e} - \mu_{e}) + \sigma_{e} \eta_{t,e}, \eta_{t,e} \sim N(0,1) \text{ iid}, 
\end{equation}  
with $\mu_e, \phi_e$ $\sigma_e$ and $s_{0,e}$ as in Section \ref{sec:bivdynmod}. For each edge $e$ in $E_i^{static}$ we have a corresponding family indicator $m_e$ and a state $s_e$ as in Section \ref{sec:bivconstmod}.
The states $s_{t,e}$ and $s_e$ are mapped to the copula parameter as in \eqref{eq:dynpar} and \eqref{eq:cpar}, i.e.
\begin{equation}
\theta_{t,e}^{m_e} = g^{-1}_{m_e}(F_Z^{-1}(s_{t,e})) \text{ and } \theta_{e}^{m_e} = g^{-1}_{m_e}(F_Z^{-1}(s_{e})), \text{ respectively}.
\end{equation}
This yields the following parameter set for a $d$-dimensional dynamic vine copula model
\begin{equation*}
\boldsymbol {\theta^{V}} = \{\mu_e, \phi_e, \sigma_e, s_{0,e}, \ldots, s_{T,e}, m_e|e \in E_i^{dyn},i=1, \ldots, d-1\} \cup \{s_{e},m_e |e \in E_i^{static},i=1, \ldots, d-1\}
\end{equation*}

Within the dynamic vine copula model, we assume that $T$ random vectors of dimension $d$, $(U_{t1}, \ldots, U_{td})_{t=1, \ldots, T}$ are generated as follows

\begin{equation}
\begin{split}
(U_{t1}, \ldots U_{td})|\boldsymbol {\theta^V} \sim &\prod_{i=1}^{d-1} \prod_{e E^{dyn}_i} c^{m_e}_{a_e,b_e;\boldsymbol {D_e}}(u_{t,a_e|\boldsymbol {D_e}}, u_{t,b_e|\boldsymbol {D_e}}; \theta_{t,e}^{m_e}) \cdot\\
& \cdot \prod_{i=1}^{d-1} \prod_{e \in E^{static}_i} c^{m_e}_{a_e,b_e;\boldsymbol {D_e}}(u_{t,a_e|\boldsymbol {D_e}}, u_{t,b_e|\boldsymbol {D_e}};\theta_e^{m_e}) ,\text{ independently}\\
\end{split}
\end{equation}
for $ t=1, \ldots, T$. Further, we assume that parameters for different edges are a priori independent and we use the same priors as specified in Sections \ref{sec:bivdynmod} and \ref{sec:bivconstmod} for the parameters of dynamic and static pair copulas, respectively. Here, the conditioned and conditioning sets $a_e, b_e$ and $\boldsymbol {D_e}$ do not depend on the time, i.e. the tree structure does not change over time.

\subsection{Sequential estimation}
\label{sec:seqest}
Since there exist $\frac{d!}{2} \cdot 2^{\binom{d-2}{2}}  $ different regular vine tree structures in $d$ dimensions (\cite{morales2010counting}), model selection is complex and it is not possible to take all possible structures into account as the dimension grows. \cite{gruber2018bayesian} estimate all trees and parameters jointly using a Bayesian approach, but their procedure is only suitable in lower dimensions and requires substantial computations. Earlier, in a frequentist setup, \cite{dissmann2013selecting} proposed a sequential selection and estimation approach for static copula parameters. This sequential approach makes model selection feasible in higher dimensions and therefore this idea is often used for vine copula based models. \cite{gruber2015sequential} employ a Bayesian approach, where they estimate parameters of static vine copula models tree by tree and fix parameters at point estimates before proceeding to the next tree. \cite{vatter2018generalized} sequentially estimate the parameters of a vine copula model, where the copula parameters are modeled by generalized additive models.
We propose a Bayesian procedure, where the copula parameters do not need to be collapsed to point estimates before proceeding to the next tree. The procedure is based on an approximation of the posterior density inspired by the frequentist sequential approach of \cite{dissmann2013selecting}. 

To simplify notation we denote by $\boldsymbol {\theta_{T_i}}$ the parameters of a dynamic vine copula corresponding to tree $T_i$ , i.e.
\begin{equation}
\boldsymbol {\theta_{T_i}} = \{\mu_e, \phi_e, \sigma_e,  s_{0,e}, \ldots,s_{T,e},  m_e, e \in E^{dyn}_i\} \cup \{s_e, m_e, e \in E^{static}_i\}.
\end{equation}
Further, we define by $U = (u_{tj})_{t=1, \ldots, T, j=1, \ldots, d}$ the data matrix and the likelihood contribution of the $i$-th tree is given by
\begin{equation}
\begin{split}
\ell_i(\boldsymbol {\theta_{T_1}}, \ldots, \boldsymbol {\theta_{T_{i}}}|U) =  &\prod_{e \in E^{dyn}_i} c^{m_e}_{a_e,b_e;\boldsymbol {D_e}}(u_{t,a_e|\boldsymbol {D_e}}, u_{t,b_e|\boldsymbol {D_e}}; \theta_{t,e}^{m_e}) \cdot \\
& \cdot \prod_{e \in E^{static}_i} c^{m_e}_{a_e,b_e;\boldsymbol {D_e}}(u_{t,a_e|\boldsymbol {D_e}}, u_{t,b_e|\boldsymbol {D_e}};\theta_e^{m_e}).
\end{split}
\end{equation}
 The complete likelihood is obtained as 
\begin{equation}
\ell(\boldsymbol {\theta_{T_1}}, \ldots, \boldsymbol {\theta_{T_{d-1}}}|U) =  \prod_{i =1}^{d-1}   \ell_i(\boldsymbol {\theta_{T_1}}, \ldots, \boldsymbol {\theta_{T_{i}}}|U),
\end{equation}
and the posterior density is proportional to

\begin{equation}
f(\boldsymbol {\theta_{T_1}}, \ldots, \boldsymbol {\theta_{T_{d-1}}}|U) \propto \prod_{i =1}^{d-1}   \ell_i(\boldsymbol {\theta_{T_1}}, \ldots, \boldsymbol {\theta_{T_{i}}}|U) \pi( \boldsymbol {\theta_{T_{i}}}).
\end{equation}
Note that the posterior density is a joint density of continuous and discrete parameters. For  discrete parameters $\boldsymbol {\delta^{disc}}$ and continuous parameters $\boldsymbol {\delta^{cont}}$ the joint density is obtained as
$$
f(\boldsymbol {\delta^{cont}}, \boldsymbol {\delta^{disc}}) = f(\boldsymbol {\delta^{cont}}|\boldsymbol {\delta^{disc}}) f(\boldsymbol {\delta^{disc}}),
$$
where $f(\boldsymbol {\delta^{cont}}|\boldsymbol {\delta^{disc}})$ is a joint probability density function and $f(\boldsymbol {\delta^{disc}})$ is a joint probability mass function.

\subsubsection*{Four-dimensional illustration for a model with static pair copulas and known tree structure}
To illustrate our idea in four dimensions, we consider only static pair copulas and the tree structure to be known. The tree structure contains the pair copulas as specified in \eqref{eq:4dimt1}, \eqref{eq:4dimt2} and \eqref{eq:4dimt3}. We assume, that we observe data $U=(u_{tj})_{t=1, \ldots, T, j=1, \ldots, 4}$.
The contributions to the likelihood corresponding to trees 1, 2 and 3 are given by

\begin{equation}
\ell_1(\boldsymbol {\theta_{T_1}}|U) = \prod_{t=1}^T c^{m_{13}}_{13}(u_{t1},u_{t3};\theta^{m_{13}}_{13}) c^{m_{23}}_{23}(u_{t2},u_{t3}; \theta^{m_{23}}_{23}) c^{m_{34}}_{34}(u_{t3},u_{t4}; \theta^{m_{34}}_{34}),
\label{eq:4dimt1}
\end{equation}

\begin{equation}
\ell_2(\boldsymbol {\theta_{T_1}}, \boldsymbol {\theta_{T_2}}|U) = \prod_{t=1}^T c^{m_{12;3}}_{12;3}(u_{t1|3},u_{t2|3}; \theta^{m_{12;3}}_{12;3}) c^{m_{24;3}}_{24;3}(u_{t2|3},u_{t4|3}; \theta^{m_{24;3}}_{24;3}) ,
\label{eq:4dimt2}
\end{equation}

\begin{equation}
\ell_3(\boldsymbol {\theta_{T_1}}, \boldsymbol {\theta_{T_2}}, \boldsymbol {\theta_{T_3}}|U) = \prod_{t=1}^T c^{m_{14;23}}_{14;23}(u_{t1|23},u_{t4|23}; \theta^{m_{14;23}}_{14;23}).
\label{eq:4dimt3}
\end{equation}

In a frequentist sequential procedure the parameters of the first tree are estimated by considering the part of the likelihood corresponding to the first tree as given in \eqref{eq:4dimt1}, ignoring the likelihood contributions of higher trees \eqref{eq:4dimt2} and \eqref{eq:4dimt3} to the parameters of the first tree. This allows to maximize $\prod_{t=1}^T c^{m_{13}}_{13}(u_{t1},u_{t3}; \theta^{m_{13}}_{13})$,  $\prod_{t=1}^T c^{m_{23}}_{23}(u_{t2},u_{t3}; \theta^{m_{23}}_{23})$ and $\prod_{t=1}^T c^{m_{34}}_{34}(u_{t3},u_{t4}; \theta^{m_{34}}_{34})$ independently. 

In the Bayesian setup the marginal posterior density of the parameters corresponding to the first tree is obtained by integrating out parameters of higher trees, i.e.
\begin{equation}
f(\boldsymbol {\theta_{T_1}}|U) \propto \ell_1(\boldsymbol {\theta_{T_1}}|U) \pi(\boldsymbol {\theta_{T_1}}) \int_{domain(\boldsymbol {\theta_{T_2}}, \boldsymbol {\theta_{T_{3}}})} \prod_{i=2}^{3} \ell_i(\boldsymbol {\theta_{T_1}},\ldots,  \boldsymbol {\theta_{T_i}}|U) \pi(\boldsymbol {\theta_{T_i}}) d \boldsymbol {\theta_{T_2}} d \boldsymbol {\theta_{T_{3}}},
\label{eq:mpost1}
\end{equation}
where $domain(\boldsymbol {\theta_{T_2}}, \boldsymbol {\theta_{T_{3}}})$ is the domain of the parameters $\boldsymbol {\theta_{T_2}}, \boldsymbol {\theta_{T_{3}}}$ (For the family indicator $m_e$ the integral is replaced by a sum). While in this illustrative example it might be possible to work with the marginal posterior \eqref{eq:mpost1}, its complexity grows fast if we consider more dimensions or allow for dynamic copulas. For example, if the second and third tree were modeled with dynamic bivariate copulas, we would need to integrate out several thousand parameters for $T=1000$. To reduce complexity, we approximate $f(\boldsymbol {\theta_{T_1}}|U)$. We make use of the following notation
$$
g(\boldsymbol \delta) \approx h(\boldsymbol \delta)
$$
to denote that a density $g$ is approximately proportional to a non negative and integrable function $h$. This means that the density $g$ is approximated by the density $h^{normalized}$ given by $h^{normalized}(\boldsymbol \delta)={ h(\boldsymbol \delta)}\left({\int_{domain(\boldsymbol \delta)} h(\boldsymbol \delta) d\boldsymbol\delta }\right)^{-1}$.

Following the idea of a frequentist sequential estimation we approximate the marginal posterior $\boldsymbol {\theta_{T_1}}|U$ by considering only the part of the likelihood corresponding to the first tree, i.e. 
\begin{equation}
\begin{split}
f(\boldsymbol {\theta_{T_1}}|U) &\approx \ell_1(\boldsymbol {\theta_{T_1}}|U) \pi(\boldsymbol {\theta_{T_1}}) = \left(\prod_{t=1}^T c^{m_{13}}_{13}(u_{t1},u_{t3}; \theta_{13}^{m_{13}}) \right) \pi(s_{13})\pi({m_{13}}) \\
& \left(\prod_{t=1}^T c^{m_{23}}_{23}(u_{t2},u_{t3}; \theta_{23}^{m_{23}}) \right) \pi(s_{23})\pi({m_{23}}) \left(\prod_{t=1}^T c^{m_{34}}_{34}(u_{t3},u_{t4}; \theta_{34}^{m_{34}}) \right) \pi(s_{34})\pi({m_{34}}).
\end{split}
\label{eq:dim4post1}
\end{equation}
This approximation simplifies sampling enormously. We do not only get rid of the integral, but in addition, the parameters corresponding to different edges are independent. To obtain samples from the posterior approximation in \eqref{eq:dim4post1} we can sample parameters of different static bivariate copula models independently by utilizing the algorithm of Section \ref{sec:bivconstmod}. In particular the parameters $m_{13},s_{13}$ are sampled from a static bivariate copula model with corresponding posterior density proportional to $\left(\prod_{t=1}^T c^{m_{13}}_{13}(u_{t1},u_{t3}; \theta^{m_{13}}) \right) \pi(s_{13})\pi({m_{13}})$.
Approximations of the posterior distribution that induce independence among parameters are commonly used in variational Bayesian approaches (\cite{wainwright2008graphical}). For example in mean field variational inference it is assumed that all parameters are independent in the posterior distribution. Our assumptions are less restrictive, since we do not assume that parameters corresponding to one pair copula are independent. Further, these parameters are updated jointly.

When estimating parameters of higher trees in a sequential frequentist procedure, we condition on estimates from lower trees. Therefore we consider now the following  density
\begin{equation}
f(\boldsymbol {\theta_{T_2}}|\boldsymbol {\theta_{T_1}}, U) \propto \ell_2(\boldsymbol {\theta_{T_1}},\boldsymbol {\theta_{T_{2}}}|U) \pi(\boldsymbol {\theta_{T_2}}) \int_{domain(\boldsymbol {\theta_{T_{3}}})} \ell_3(\boldsymbol {\theta_{T_1}},\boldsymbol {\theta_{T_{2}}} ,  \boldsymbol {\theta_{T_{3}}}|U) \pi(\boldsymbol {\theta_{T_3}}) d \boldsymbol {\theta_{T_3}}.
\end{equation}
We utilize a similar approximation as in \eqref{eq:dim4post1} and obtain
\begin{equation}
\begin{split}
f(\boldsymbol {\theta_{T_2}}|\boldsymbol {\theta_{T_1}}, U) \approx \ell_2(\boldsymbol {\theta_{T_1}},\boldsymbol {\theta_{T_{2}}}|U) \pi(\boldsymbol {\theta_{T_2}})= 
& \left(\prod_{t=1}^T c^{m_{12;3}}_{12;3}(u_{t1|3},u_{t2|3}; \theta_{12;3}^{m_{12;3}})\right) \pi(s_{12;3})\pi(m_{12;3}) \\
&\left( \prod_{t=1}^T  c^{m_{24;3}}_{24;3}(u_{t2|3},u_{t4|3}; \theta_{24;3}^{m_{24;3}}) \right) \pi(s_{24;3})\pi(m_{24;3}) .
\end{split}
\end{equation}
The pseudo data $u_{t1|3}=h_{1|3}(u_{t1}|u_{t3};\theta^	{m_{13}}_{13},m_{13})=\frac{d}{du_3}C^{m_{13}}_{13}(u_{t1},u_{3};\theta_{13}^{m_{13}})\Big|_{u_3=u_{t3}}$ , 
$u_{t2|3}=$ \newline $h_{2|3}(u_{t2}|u_{t3};\theta^{m_{23}}_{23},m_{23})$ and $u_{t4|3}=h_{4|3}(u_{t4}|u_{t3}, \theta^{m_{34}}_{34},m_{34}), t=1, \ldots, T$ only depend on parameters of the first tree, on which we condition on. Further, the posterior density factorizes as in \eqref{eq:dim4post1} and we can sample parameters corresponding to different edges independently. In particular $s_{12;3}, m_{12;3}$ are sampled from a static bivariate copula model with posterior density proportional to $\left(\prod_{t=1}^T c^{m_{12;3}}_{12;3}(u_{t1|3},u_{t2|3}; \theta_{12;3}^{m_{12;3}})\right) \pi(s_{12;3})\pi(m_{12;3})$, where $\{u_{t1|3}, u_{t2|3}, t=1, \ldots, T\}$ is interpreted as observed data.
For the third tree we obtain
\begin{equation}
\begin{split}
f(\boldsymbol {\theta_{T_3}}|\boldsymbol {\theta_{T_1}},\boldsymbol {\theta_{T_2}}, U) &\propto \ell_3(\boldsymbol {\theta_{T_1}},\boldsymbol {\theta_{T_{2}}},\boldsymbol {\theta_{T_{3}}}|U) \pi(\boldsymbol {\theta_{T_3}}) \\
&= \left(\prod_{t=1}^T c^{m_{14;23}}_{14;23}(u_{t1|23},u_{t4|23}; \theta_{14;23}^{m_{14;23}})\right) \pi(s_{14;23})\pi(m_{14;23})
\end{split}.
\label{eq:4dimt3_full}
\end{equation}

As before, $u_{t1|23}$ and $u_{t4|23}$ only depend on parameters from lower trees, on which we condition on. Interpreting $\{u_{t1|23},u_{t4|23}, t=1, \ldots, T\}$ as observed data, \eqref{eq:4dimt3_full} is the posterior density of a static bivariate copula model as introduced in Section \ref{sec:bivconstmod}.

To obtain samples from the posterior density $f(\boldsymbol {\theta_{T_1}},\boldsymbol {\theta_{T_2}},\boldsymbol {\theta_{T_3}}|U)$, we utilize the approximations above to first sample $\boldsymbol {\theta_{T_1}}$ from $f(\boldsymbol {\theta_{T_1}}|U)$, then $\boldsymbol {\theta_{T_2}}$ from  $f(\boldsymbol {\theta_{T_2}}|\boldsymbol {\theta_{T_1}},U)$ and then   $\boldsymbol {\theta_{T_3}}$ from $f(\boldsymbol {\theta_{T_3}}|\boldsymbol {\theta_{T_2}},\boldsymbol {\theta_{T_1}},U)$, i.e. we employ a collapsed Gibbs sampler (\cite{liu1994collapsed}). To update the parameters we use the sampling procedure of Section \ref{sec:bivconstmod}.  This sampler simulates a Markov chain, where subsequent draws are autocorrelated. So, by applying this sampler we obtain a sample of $\boldsymbol {\theta_{T_i}}$ in the $r$-th iteration, denoted by $\boldsymbol {\theta_{T_i}^r}$, which depends on the previous value  $\boldsymbol {\theta_{T_i}^{r-1}}$. While this is not a problem for conventional Gibbs samplers, as in Metropolis-Hastings within Gibbs, this can lead to undesired samples for collapsed Gibbs schemes as shown by \cite{van2015metropolis}. Following \cite{van2015metropolis}, we can circumvent this problem by running the updates for $\boldsymbol {\theta_{T_i}}$ with starting value $\boldsymbol {\theta_{T_i}^{r-1}}$ for $k$ iterations. We set $\boldsymbol {\theta_{T_i}^r}$ equal to the update obtained in the $k$-th step. If we choose $k$ large enough, $\boldsymbol {\theta_{T_i}^r}$ will be almost independent of $\boldsymbol {\theta_{T_i}^{r-1}}$. Thus, in total, $R \cdot k$ draws are obtained and $R$ draws are stored for each parameter. We obtain the following procedure

\vspace*{0.2cm}
\noindent
-- Set starting values $\boldsymbol {\theta_{T_{1}}^0}$, $\boldsymbol {\theta_{T_{2}}^0}$, $\boldsymbol {\theta_{T_{3}}^0}$

\vspace*{0.2cm}
\noindent
-- For $r=1, \ldots, R$ do

\vspace*{0.2cm}
\noindent
\hspace*{0.2cm}-- For $i= 1, \ldots, 3$ do

\begin{itemize}
\item[--] Use the sampler of Section \ref{sec:bivconstmod} to sample from
 $f(\boldsymbol {\theta_{T_1}}|U)$ if $i=1$, from $f(\boldsymbol {\theta_{T_2}}|\boldsymbol {\theta_{T_1}^r},U)$ if $i=2$ or from  $f(\boldsymbol {\theta_{T_3}}|\boldsymbol {\theta_{T_1}^r},\boldsymbol {\theta_{T_2}^r},U)$ if $i=3$. The sampler is run for $k$ iterations using $\boldsymbol {\theta_{T_{i}}^{r-1}}$ as starting value. We set $\boldsymbol {\theta_{T_{i}}^{r}}$ equal to the sample obtained in the $k$-th iteration.
\item[--] The pseudo data for the next tree is constructed utilizing the $h$ functions defined in \eqref{eq:hdef}.

For $i=1$ the pseudo data is determined as $u_{t1|3}^r=h_{1|3}(u_{t1}|u_{t3};(\theta^{m_{13}}_{13})^r,m_{13}^r)$,
$u_{t2|3}^r=h_{2|3}(u_{t2}|u_{t3};(\theta^{m_{23}}_{23})^r,m_{23}^r)$ and $u_{t4|3}^r=h_{4|3}(u_{t4}|u_{t3}; (\theta^{m_{34}}_{34})^r,m_{34}^r), t=1, \ldots, T$. 

For $i=2$ the pseudo data is determined as  $u_{t1|23}^r=$  $h_{1|2;3}(u_{t1|3}^r,u_{t2|3}^r;(\theta^{m_{12;3}}_{12;3})^r,m_{12;3}^r)$ and $u_{t4|23}^r = h_{4|2;3}(u_{t4|3}^r,u_{t2|3}^r;(\theta_{24;3}^{m_{24;3}})^r,m_{24;3}^r)$ $, t=1, \ldots, T.$
\end{itemize}

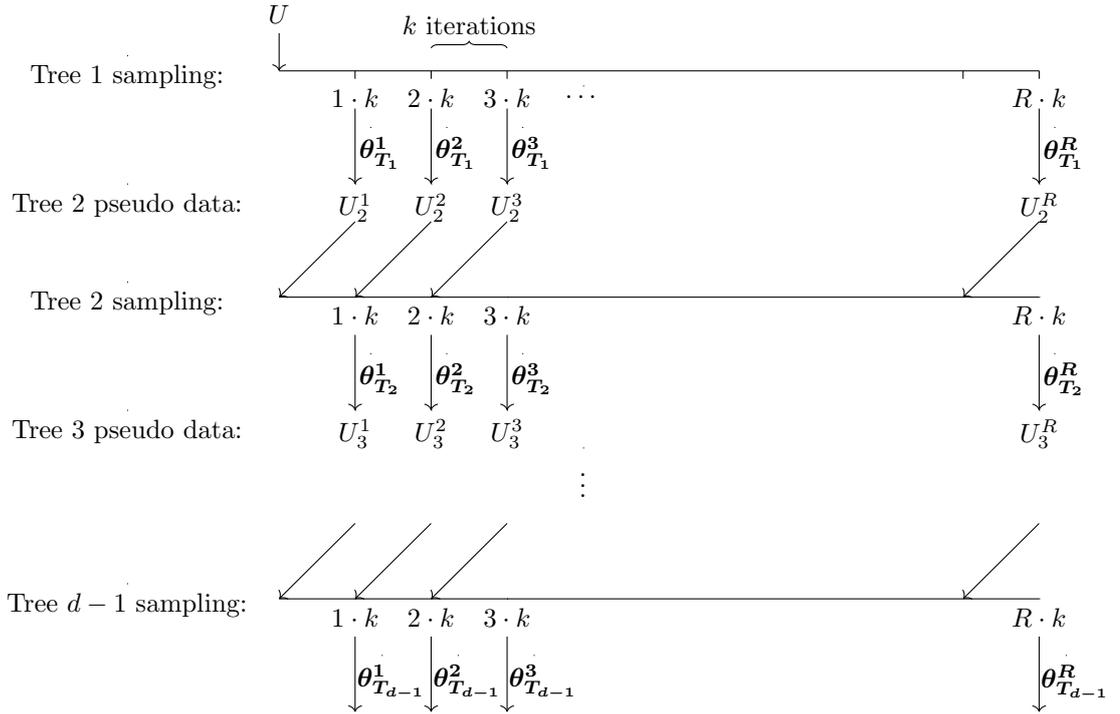
\begin{figure}[H]

 \begin{tikzpicture}[decoration=brace]
     \draw(2,1)--(2,1) node[below] {$\dU$};
              \draw [->] (2,0.5) to node[draw=none, fill = none, above] {} (2,0);

    \draw(2,0)--(12,0);
     \draw(0,0.2)--(0,0.2) node[below] {Tree 1 sampling:};
     \draw(3,0)--(3,-3pt) node[below] {$1 \cdot k$};
     \draw(4,0)--(4,-3pt) node[below] {$2 \cdot k$};
     \draw(5,0)--(5,-3pt) node[below] {$3 \cdot k$};
     \draw(6,-0.2)--(6,-0.2) node[below] {\ldots};
          \draw(11,0)--(11,-3pt) node[below] {};
     \draw(12,0)--(12,-3pt) node[below] {$R \cdot k$};
         \draw[decorate, yshift=2ex]  (4,0) -- node[above=0.4ex] {$k$ iterations}  (5,0);

     
      \draw(0,-1.5)--(0,-1.5) node[below] {Tree 2 pseudo data:};
     \draw(3,-1.5)--(3,-1.5) node[below] {$\dU_2^1$};
     \draw(4,-1.5)--(4,-1.5) node[below] {$\dU_2^2$};
     \draw(5,-1.5)--(5,-1.5) node[below] {$\dU_2^3$};
     \draw(12,-1.5)--(12,-1.5) node[below] {$\dU_2^R$};

          \draw [->] (3,-0.5) to node[draw=none, fill = none, left] {} (3,-1.5);
          \draw [->] (4,-0.5) to node[draw=none, fill = none, left] {} (4,-1.5);
         \draw [->] (5,-0.5) to node[draw=none, fill = none, left] {} (5,-1.5);
         \draw [->] (12,-0.5) to node[draw=none, fill = none, left] {} (12,-1.5);

     \draw(3.2,-0.75)--(3.2,-0.75) node[below] {$\hspace*{0.25cm}\boldsymbol{\theta_{T_1}^1}$};
     \draw(4.2,-0.75)--(4.2,-0.75) node[below] {$\hspace*{0.25cm}\boldsymbol{\theta_{T_1}^2}$};
     \draw(5.2,-0.75)--(5.2,-0.75) node[below] {$\hspace*{0.25cm}\boldsymbol{\theta_{T_1}^3}$};
     \draw(12.2,-0.75)--(12.2,-0.75) node[below] {$\hspace*{0.25cm}\boldsymbol{\theta_{T_1}^R}$};

    \draw(2,-3)--(12,-3);
     \draw(0,-2.8)--(0,-2.8) node[below] {Tree 2 sampling:};
     \draw(3,-3)--(3,-3) node[below] {$1\cdot k$};
     \draw(4,-3)--(4,-3) node[below] {$2\cdot k$};
     \draw(5,-3)--(5,-3) node[below] {$3\cdot k$};
     \draw(11,-3)--(11,-3) node[below] {};
     \draw(12,-3)--(12,-3) node[below] {$R \cdot k$};
 
         \draw [->] (3,-2) to node[draw=none, fill = none, above] {} (2,-3);
          \draw [->] (4,-2) to node[draw=none, fill = none, above] {} (3,-3);
          \draw [->] (5,-2) to node[draw=none, fill = none, above] {} (4,-3);
          \draw [->] (12,-2) to node[draw=none, fill = none, above] {} (11,-3);

     \draw(0,-4.5)--(0,-4.5) node[below] {Tree 3 pseudo data:};
     \draw(3,-4.5)--(3,-4.5) node[below] {$\dU_3^1$};
     \draw(4,-4.5)--(4,-4.5) node[below] {$\dU_3^2$};
     \draw(5,-4.5)--(5,-4.5) node[below] {$\dU_3^3$};
     \draw(12,-4.5)--(12,-4.5) node[below] {$\dU_3^R$};

          \draw [->] (3,-3.5) to node[draw=none, fill = none, above] {} (3,-4.5);
          \draw [->] (4,-3.5) to node[draw=none, fill = none, above] {} (4,-4.5);
         \draw [->] (5,-3.5) to node[draw=none, fill = none, above] {} (5,-4.5);
         \draw [->] (12,-3.5) to node[draw=none, fill = none, above] {} (12,-4.5);

       \draw(3.2,-3.75)--(3.2,-3.75) node[below] {$\hspace*{0.25cm}\boldsymbol{\theta_{T_2}^1}$};
     \draw(4.2,-3.75)--(4.2,-3.75) node[below] {$\hspace*{0.25cm}\boldsymbol{\theta_{T_2}^2}$};
     \draw(5.2,-3.75)--(5.2,-3.75) node[below] {$\hspace*{0.25cm}\boldsymbol{\theta_{T_2}^3}$};
     \draw(12.2,-3.75)--(12.2,-3.75) node[below] {$\hspace*{0.25cm}\boldsymbol{\theta_{T_2}^R}$};

  \draw(6,-5)--(6,-5) node[below] {\vdots };

    \draw(2,-7)--(12,-7);
     \draw(0,-6.8)--(0,-6.8) node[below] {Tree $d-1$ sampling:};
     \draw(3,-7)--(3,-7) node[below] {$1 \cdot k$};
     \draw(4,-7)--(4,-7) node[below] {$2 \cdot k$};
     \draw(5,-7)--(5,-7) node[below] {$3 \cdot k$};
     \draw(12,-7)--(12,-7) node[below] {$R \cdot k$};

     \draw [->] (3,-6) to node[draw=none, fill = none, above] {} (2,-7);
          \draw [->] (4,-6) to node[draw=none, fill = none, above] {} (3,-7);
         \draw [->] (5,-6) to node[draw=none, fill = none, above] {} (4,-7);
         \draw [->] (12,-6) to node[draw=none, fill = none, above] {} (11,-7);

            \draw [->] (3,-7.5) to node[draw=none, fill = none, above] {} (3,-8.5);
          \draw [->] (4,-7.5) to node[draw=none, fill = none, above] {} (4,-8.5);
         \draw [->] (5,-7.5) to node[draw=none, fill = none, above] {} (5,-8.5);
         \draw [->] (12,-7.5) to node[draw=none, fill = none, above] {} (12,-8.5);

      \draw(3.35,-7.75)--(3.35,-7.75) node[below] {$\hspace*{0.25cm}\boldsymbol{\theta_{T_{d-1}}^1}$};
     \draw(4.35,-7.75)--(4.35,-7.75) node[below] {$\hspace*{0.25cm}\boldsymbol{\theta_{T_{d-1}}^2}$};
     \draw(5.35,-7.75)--(5.35,-7.75) node[below] {$\hspace*{0.25cm}\boldsymbol{\theta_{T_{d-1}}^3}$};
     \draw(12.35,-7.75)--(12.35,-7.75) node[below] {$\hspace*{0.25cm}\boldsymbol{\theta_{T_{d-1}}^R}$};






  \end{tikzpicture}
\caption{Graphical representation of the proposed sampler without selection of the type of dependence and without structure selection. Here $U \in [0,1]^{T \times d}$ denotes the data matrix used for fitting the model and {$\protect\dU_l^r$ } denotes the pseudo data for tree $l$ obtained from parameter draws of the previous tree (tree $l-1$) in iteration $r$.}
\label{fig:tik_algo}
\end{figure}

In this procedure, no point estimates of the copula parameters $\boldsymbol {\theta_1}, \boldsymbol {\theta_2}, \boldsymbol {\theta_3}$ are required. 
We can further extend the procedure in the following ways.
\begin{itemize}
\item a) The loops over $i$ and $r$ can be exchanged. We can first obtain $R$ samples from the first tree, then obtain $R$ samples from the second tree and then all $R$ samples from the third tree. This is visualized in Figure \ref{fig:tik_algo}. If the tree structure was not known we could select the tree structure of the first tree, then obtain $R$ samples from the parameters of the first tree. Based on these samples, we can construct the pseudo data, which can be used to select the tree structure of the second tree and so on. Based on the (pseudo) data of a certain tree level, the corresponding structure can be selected as a maximum spanning tree. This is similar to the algorithm of \cite{dissmann2013selecting}.
\item b) The parameters of different edges of a tree are sampled independently by utilizing the sampler of Section \ref{sec:bivconstmod} for the static copula model. If we did not know that Kendall's $\tau$ was static we could in addition run the sampler of Section \ref{sec:bivdynmod} for the dynamic bivariate copula model. We can decide between the dynamic, the static and the independence model as outlined in Section \ref{sec:WAIC}. Here it is important that these decisions for the type of dependence can be made independently for each edge of the tree.
\end{itemize}

\subsubsection*{The general procedure in $d$ dimensions with  vine structure selection}

Based on the four-dimensional illustration, we now formulate our procedure for a $d$-dimensional dynamic vine copula as introduced in Section \ref{sec:vinemodelspec}, incorporating extensions a) and b). The tree structure and the sets $E^{dyn}_i, E^{static}_i, E^{ind}_i$ are selected sequentially as we move up the trees and are fixed at point estimates. In \cite{gruber2018bayesian} searching among different structures within a full Bayesian procedure resulted in very long computation times for static copula models. Here it would be even worse, since we deal with more complex dynamic pair copulas. 
Note that both, the sets $E^{dyn}_i, E^{static}_i, E^{ind}_i$ and the tree structure do not change over time.

We propose the following approach with iterations parameter $R$, burn-in parameter $burnin$ and thinning parameter $k$ for structure selection and parameter estimation. Note that, as mentioned above, $R \cdot k$ draws are obtained in total and $R$ iterations are stored for each parameter.
\begin{enumerate}
\item Select the tree structure of tree $T_1$:
For all edges $e$ that are allowed in the first tree $T_1$, i.e. for all pairs $(a_e,b_e)$ with $1\leq a_e <  b_e \leq d$, estimate $\tau_{a_e,b_e}$ by the empirical Kendall's $\tau$ using $\{u_{t,a_e}, u_{t,b_e}, t=1, \ldots, T\}$. The structure of tree $T_1$ is selected as the maximum spanning tree among those edges, where the absolute value of empirical Kendall's $\tau$ serves as the corresponding weight.

\item \begin{enumerate}
\item For each edge $e \in E_1$ in tree $T_1$, with corresponding observations $\{u_{t,a_e}, u_{t,b_e}, t=1, \ldots, T\}$, run the samplers of Sections \ref{sec:bivdynmod} and \ref{sec:bivconstmod} for the bivariate dynamic and static copula models. The samplers are run for $R\cdot k$ iterations and we thin the samples with factor $k$. 
\item For each edge $e \in E_1$, we select among the bivariate dynamic, static and the bivariate independence copula model as discussed in Section \ref{sec:WAIC}.
\item  For each edge $e \in E_1$, the pseudo data for the next tree is obtained as
\begin{equation}
\begin{split}
u_{t,a_e|b_e }^r ={h}_{a_e|b_e}(u_{t,a_e}|u_{t,b_e}; (\theta_{t,e}^{m_e})^r, m_e^r), \\
u_{t,b_e|a_e}^r = {h}_{b_e|a_e}(u_{t,b_e}|u_{t,a_e}; (\theta_{t,e}^{m_e})^r, m_e^r),
\end{split}
\label{eq:pseudodata1}
\end{equation}
for $r=1, \ldots, R, t=1, \ldots, T$, if the dynamic copula was selected for edge $e$. If the static copula was selected we replace $\theta_{t,e}^{m_e}$ by $\theta_e^{m_e}$ in \eqref{eq:pseudodata1}. For the independence copula model we use $u_{t,a_e|b_e }^r=u_{t,a_e}$ and $u_{t,b_e|a_e }^r=u_{t,b_e}$.
\end{enumerate}

\item Set $l=2$.

\item Select the tree structure of tree $T_l$:
For all edges that are allowed in tree $T_l$ according to the proximity condition, estimate Kendall's $\tau$ of edge $e = (a_e,b_e;\boldsymbol {D_e})$ denoted by $\tau_{a_e,b_e;\boldsymbol {D_e}}$ by the empirical Kendall's $\tau$. Therefore we use posterior mode estimates of the pseudo data
\{$\hat u_{t,a_e|\boldsymbol {D_e}}, \hat u_{t,b_e|\boldsymbol {D_e}}, t=1, \ldots, T$\}, where
 $ \hat u_{t,a_e|\boldsymbol {D_e}}$ is the mode of the univariate kernel density estimate of $\{u_{t,a_e|\boldsymbol {D_e}}^r, r=burnin+1, \ldots, R\}$ and $\hat u_{t,b_e|\boldsymbol {D_e}}$ is obtained similarly. Here the posterior mode pseudo data $\{\hat u_{t,a_e|\boldsymbol {D_e}}, \hat u_{t,b_e|\boldsymbol {D_e}}, t=1, \ldots, T\}$ are treated as an i.i.d. sample for the estimation of $\tau_{a_e,b_e;\boldsymbol {D_e}}$. The structure of tree $T_l$ is selected as the maximum spanning tree among those edges, where the absolute value of empirical Kendall's $\tau$ serves as the corresponding weight.

\item \begin{enumerate} 
\item For each edge $e \in E_l$ in tree $T_l$, with corresponding pseudo data
$\{u_{t,a_e|\boldsymbol {D_e}}^{r}$, $u_{t,b_e|\boldsymbol {D_e}}^{r}$, $t=1, \ldots, T, r=1, \ldots, R\}$, obtain $R$ samples (based on a total of $R\cdot k$ MCMC draws) from the bivariate dynamic and static copula models utilizing the approaches of Sections \ref{sec:bivdynmod} and \ref{sec:bivconstmod}.
For the static bivariate copula we proceed as follows for an edge $e$.
\begin{itemize}
\item[--] Set starting values $s_e^0, m_e^0$.

\item[--] For $r=1, \ldots, R$: obtain $k$ samples of $s_e, m_e$ from a static bivariate copula model based on data $\{u_{t,a_e|\boldsymbol {D_e}}^{r}$, $u_{t,b_e|\boldsymbol {D_e}}^{r}$, $t=1, \ldots, T\}$. We use $s_e^{r-1}, m_e^{r-1}$ as starting value and set $s_e^{r}, m_e^{r}$ to the sample obtained in the $k$-th iteration.
\end{itemize}
For the dynamic copula model we proceed similarly.

\item We select for each edge $e \in E_l$ among the bivariate dynamic, static and the bivariate independence copula model as explained in Section \ref{sec:WAIC}.
 \item For each edge $e \in E_l$, the pseudo data for the next tree is obtained as
\begin{equation}
\begin{split}
u_{t,a_e|b_e \cup \boldsymbol {D_e}}^r = {h}_{a_e|b_e;\boldsymbol {D_e}}(u_{t,a_e|\boldsymbol {D_e}}^r|u_{t,b_e|\boldsymbol {D_e}}^r; (\theta_{t,e}^{m_e})^r, m_e^r), \\
u_{t,b_e|a_e \cup \boldsymbol {D_e}}^r = {h}_{b_e|a_e;\boldsymbol {D_e}}(u_{t,b_e|\boldsymbol {D_e}}^r|u_{t,a_e|\boldsymbol {D_e}}^r; (\theta_{t,e}^{m_e})^r, m_e^r),
\end{split}
\label{eq:pseudodata2}
\end{equation}
for $r=1, \ldots, R, t=1, \ldots, T$, if the dynamic copula was selected for edge $e$. If the static copula was selected we replace $\theta_{t,e}^{m_e}$ by $\theta_e^{m_e}$ in \eqref{eq:pseudodata2}. For the independence copula model we set $u_{t,a_e|b_e\cup \boldsymbol {D_e} }^r=u_{t,a_e|\boldsymbol {D_e}}^r$ and $u_{t,b_e|a_e \cup \boldsymbol {D_e} }^r=u_{t,b_e|\boldsymbol {D_e}}^r$.
\end{enumerate}

\item If $l < d-1$, set $l=l+1$ and go to 4.
\end{enumerate}
\subsubsection*{Runtime and scalability}
The MCMC samplers in step 2.(a)  and the ones in step 5.(a) can be run in parallel, respectively. The MCMC samplers are the main drivers for the runtime and therefore parallelization speeds up computation a lot. When enough cores, i.e. at least $d-1$ cores for $d$-dimensional data, are available, we observed that the computation time for one tree is no more than 40 minutes for time series data of length $T = 1000$ with $R=1100, burnin=100, k=25$, independently of $d$. For estimating a full vine (i.e. a vine without truncation), we expect that the computation time grows roughly linearly with the dimension $d$ and a full vine in 11 dimension (containing 10 trees) should take no more than $10 \cdot 40$ minutes. But in higher dimensions it is often not necessary to estimate all trees, e.g. we expect the runtime for a 100-dimensional vine, truncated after the 10-th tree, to be not much more than $10 \cdot 40$ minutes. Thus, in combination with truncation, we expect our method to scale very well to higher dimensions.

\subsection{Simulation study}
\label{sec:simstudy2}
With this simulation study, we aim to obtain a first impression of the ability of the procedure proposed in Section \ref{sec:seqest} to recover trajectories of Kendall's $\tau$ and of the ability to select copula families and the type of dependence (dynamic, static, zero).
A more extensive simulation study to investigate the potential of the novel approach in more detail is planned for the future.

First, we assume the tree structure to be known and the steps for the vine structure selection in our procedure are left out. 
This allows to compare the true and estimated Kendall's $\tau$ values for each pair copula.  Afterwards we allow for vine structure selection. In this case our procedure might select different tree structures. Thus the true trajectories of Kendall's $\tau$ and the copula family for some pair copulas included in the selected vine structure may not be directly known. We deal with this case by comparing simulations from the true and the estimated model. In addition we compare average log likelihoods of true and estimated models.

\subsubsection*{Known tree structure}

We consider the tree structure presented in Figure \ref{fig:treestruct}. The corresponding families are chosen from the set: $\{ $Independence, Gaussian, Student t(df=2), Student t(df=4), Student t(df=8), eGumbel, eClayton$\}$. For each pair copula we simulate one trajectory of length $T=1000$ for Kendall's $\tau$ from an AR(1) process. The chosen copula families and the parameters of the AR(1) processes are specified in Appendix \ref{sec:app_sim}. 
We keep the tree structure, the choice of the families and the trajectories for Kendall's $\tau$ fixed and simulate 100 times from this model.

\begin{figure}[H]
\centerline{%
\includegraphics[trim={0 0cm 0 0},width=1.0\textwidth]{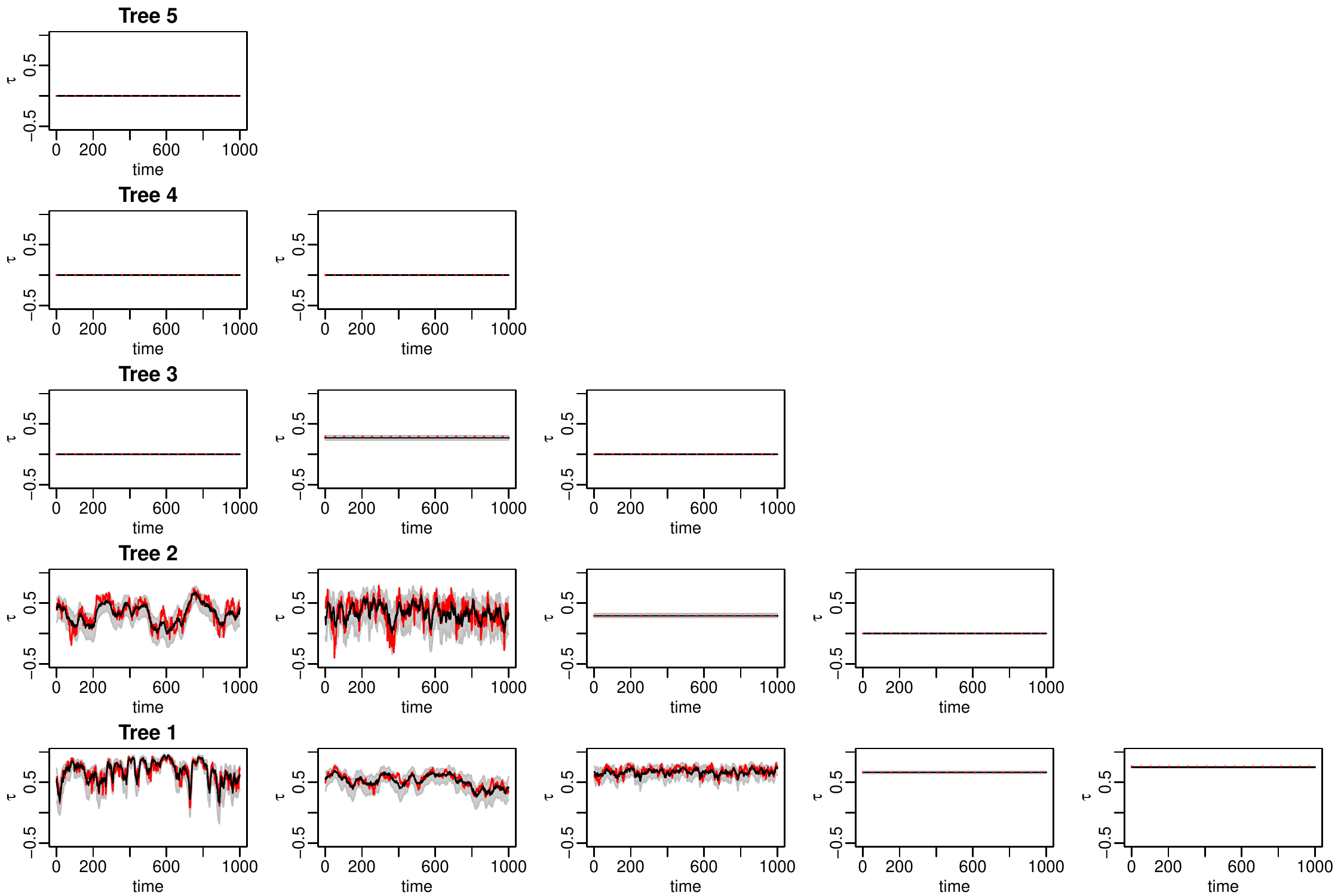}%
}%
\caption{This plot corresponds to a fitted model for one simulated data set. Posterior mode estimates of Kendall's $\tau$ at time $t$ are plotted against $t$ for each pair copula (black lines). The posterior mode estimates are obtained from marginal (univariate) kernel density estimates of the corresponding Kendall's $\tau$ parameter. A $90\%$ credible region constructed from the estimated $5\%$ and $95\%$ posterior quantiles is added in grey. True values of Kendall's $\tau$ are added in red. 
}
\label{fig:sim1_run3}
\end{figure}

\begin{figure}[H]
\centerline{%
\includegraphics[trim={0 0cm 0 0},width=1.0\textwidth]{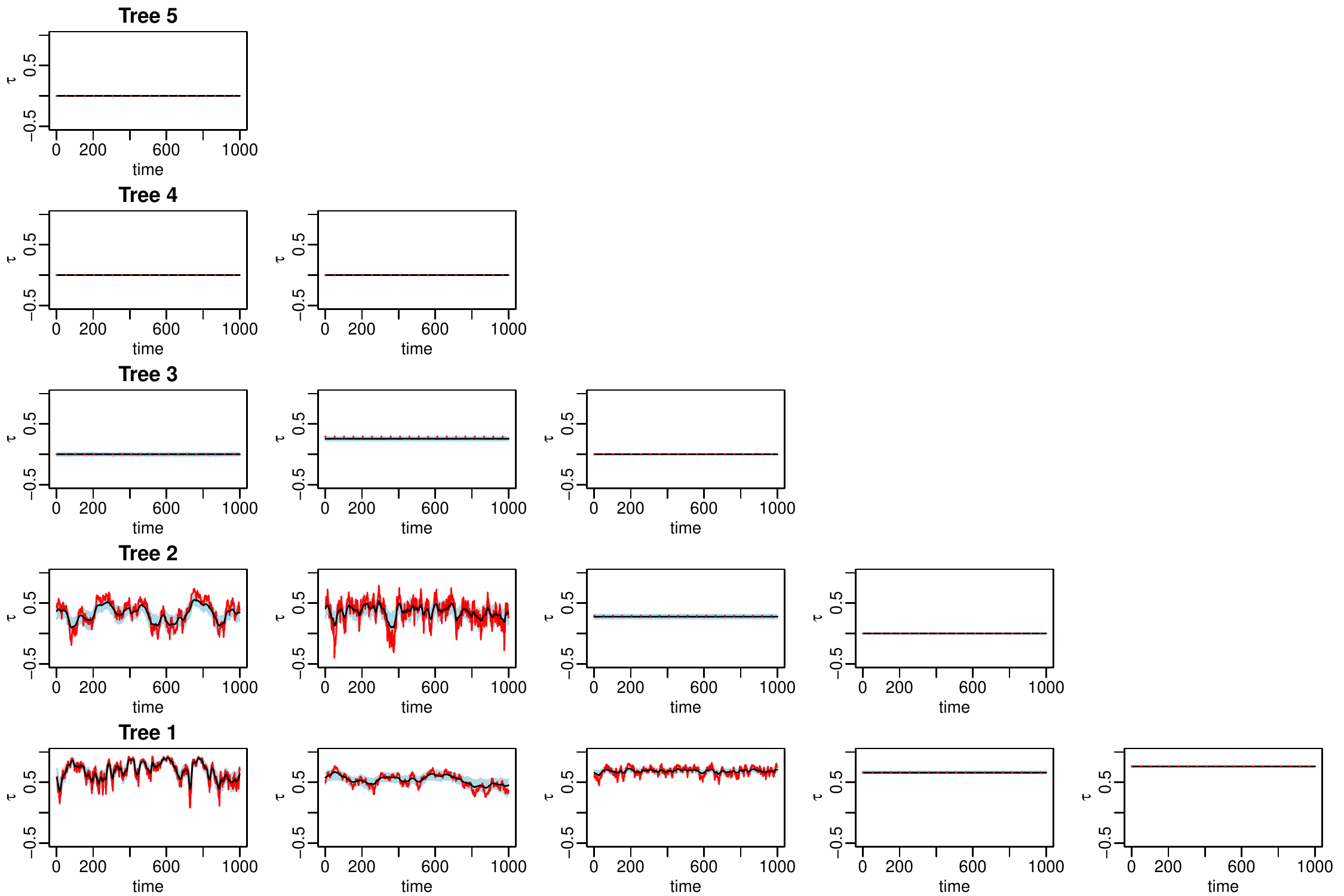}%
}%
\caption{In this plot we consider 100 estimated models. The mean of 100 posterior mode estimates of Kendall's $\tau$ at time $t$ is plotted against $t$ for each pair copula (black line). The posterior mode estimates are obtained from marginal (univariate) kernel density estimates of the corresponding Kendall's $\tau$ parameter. The blue region is constructed from the empirical $5\%$ and $95\%$ quantiles of the 100 posterior mode estimates. True values of Kendall's $\tau$ are added in red. 
}
\label{fig:sim1_dyn}
\end{figure}

 For each of the 100 simulated data sets we run the algorithm proposed in Section \ref{sec:seqest}. Within our sequential procedure, we set $R=1100$, $k=25$ and $burnin=100$. This means that within the procedure $1100 \cdot 25$ draws have been obtained for each parameter, whereas $1100$ iterations are stored. Of these $1100$ stored iterations the first $100$ are discarded for burn-in.
From Table \ref{tab:sim2_fam} we see that for each pair copula in the first two trees the correct family was selected in at least 94 out of 100 cases. In Tree 3, the two independence copulas were detected in 69 and 100 out of 100 cases. The static copula in Tree 3 was detected in 88 out of 100 cases.  The independence copulas in Trees 4 and 5 were detected every time. Here we count a Student t copula as correctly detected if it was selected as a Student t copula, independently of the degrees of freedom parameter.
Table \ref{tab:sim2_fam} also shows how often the correct type of dependence (dynamic, static, zero) was selected.
For one pair copula in the first tree the correct type was only detected in 75 out of 100 cases. The corresponding Kendall's $\tau$ is shown in the fifth row, third column in Figure \ref{fig:sim1_run3}. We see that this Kendall's $\tau$ does not change a lot over time. So it is difficult to distinguish between the dynamic and the static model for this pair copula. Except for this pair copula, the type of dependence of pair copulas in the first two trees was detected in at least 93 out of 100 cases. In trees, higher than tree 2, the correct type was selected in at least 69 out of 100 cases. We think that these are reasonable results for the selection of the family and of the type of dependence for the pair copulas, that make up the dynamic vine copula. In addition, Figures \ref{fig:sim1_run3} and \ref{fig:sim1_dyn} illustrate that our procedure can recover the simulated trajectories of Kendall's $\tau$. In these figures, we show marginal (univariate) posterior mode estimates of Kendall's $\tau$ parameters, which will be utilized later (Section \ref{sec:app}) as point estimates.

\begin{table}[H]
\centering
\begin{tabular}{c|rrrrr|rrrrr}
Tree & \multicolumn{5}{c}{Copula family}& \multicolumn{5}{c}{Type of dependence}\\
\hline
 5 & 100 &  &  &  &   & 100 &  &  &  &    \\ 
 4 & 100 & 100 &  &  &   & 100 & 100 &  &  & \\ 
 3 & 69 & 88 & 100 &  &   & 69 & 78 & 100 &  & \\ 
 2 & 100 & 100 & 94 & 100 &   & 100 & 96 & 97 & 100 &  \\ 
 1 & 94 & 100 & 100 & 99 & 100  & 100 & 93 & 75 & 98 & 97\\ 
   \hline
\end{tabular}
\caption{This table shows how often the correct copula family and how often the correct type of dependence (dynamic, static, zero) was selected out of the 100 simulations for each pair copula of the dynamic vine copula. There are $6-i$ pair copulas in the $i$-th tree. The selected copula family for an edge $e$ is the marginal posterior mode estimate of $m_e$, i.e. the family that occurs most frequently among the posterior samples for $m_e$. } 
\label{tab:sim2_fam}
\end{table}

\subsubsection*{Unknown tree structure}
\begin{figure}[H]
\centerline{%
\includegraphics[trim={0 6cm 0 0},width=1.0\textwidth]{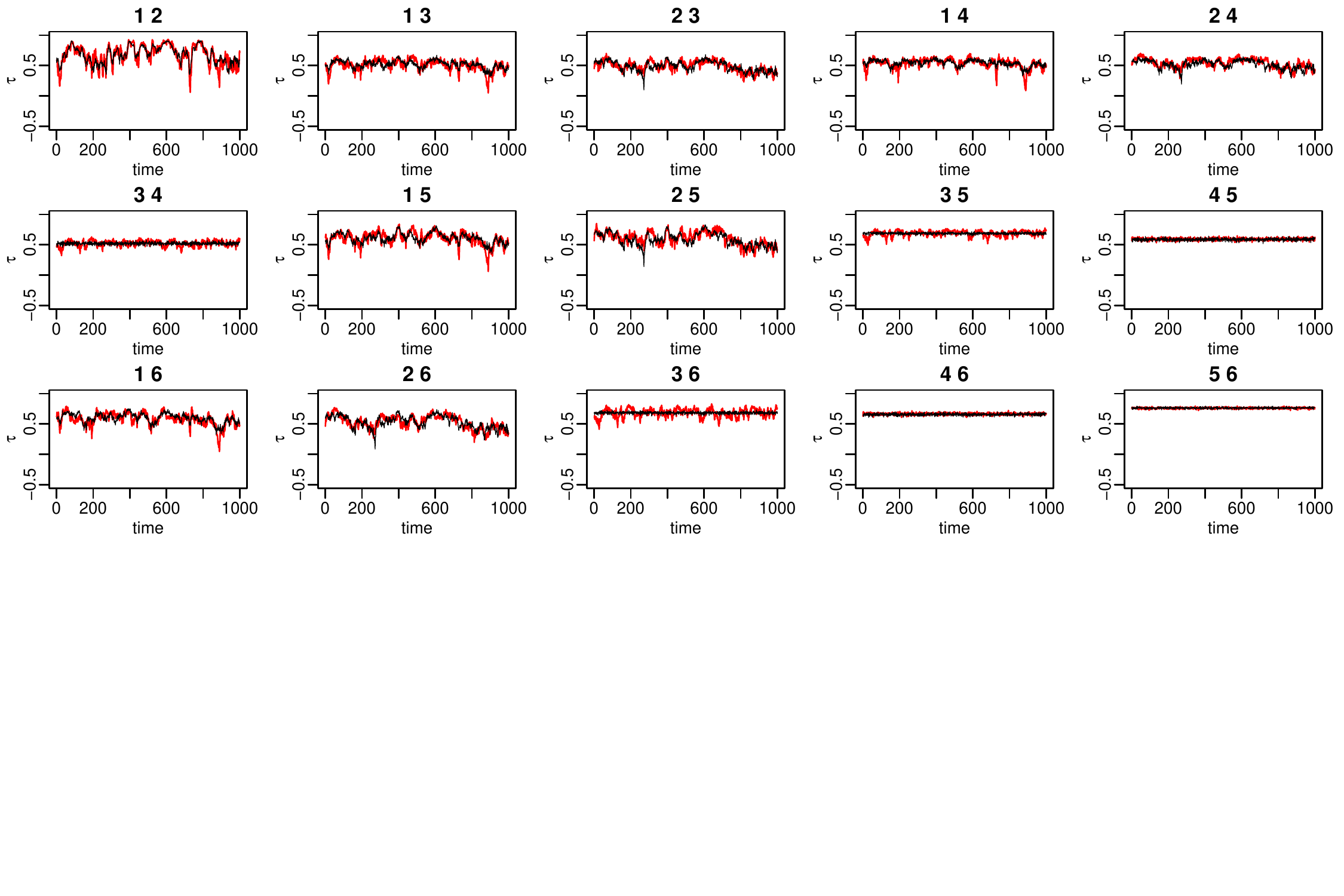}%
}%
\caption{This plot corresponds to a fitted model for one simulated data set. The empirical unconditional Kendall's $\tau$ estimate at time $t$, obtained from simulations from the fitted model, is plotted against $t$ for each pair $(U_i,U_j), i,j \in \{1, \ldots, 6\},i<j$ (black line). True Kendall's $\tau$ values determined by simulating from the true model 500 times are added in red.}
\label{fig:sim_us_run1}
\end{figure}

\begin{figure}[H]
\centerline{%
\includegraphics[trim={0 6cm 0 0},width=1.0\textwidth]{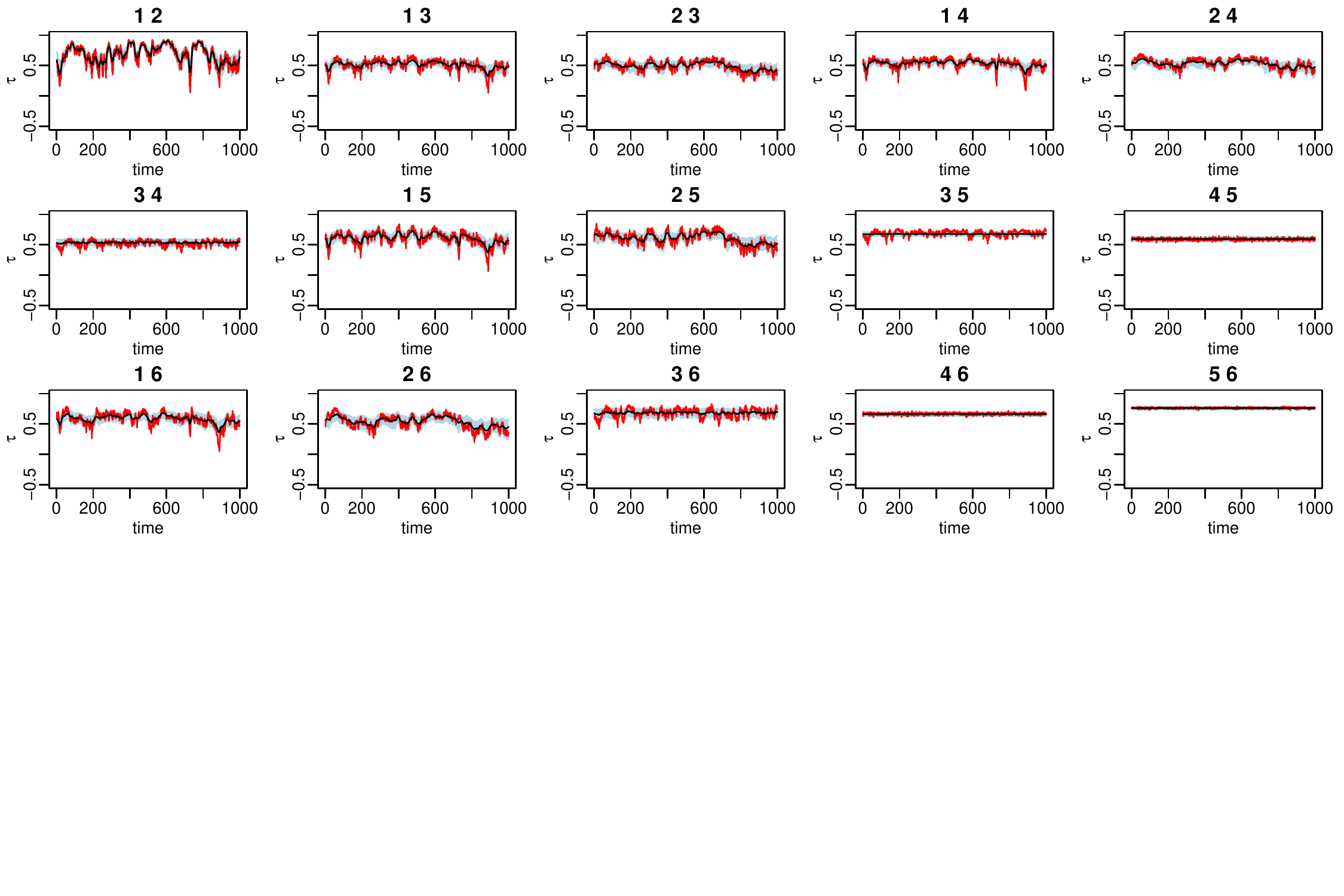}%
}%
\caption{In this plot we consider 100 fitted models. The mean of 100 empirical estimates of the unconditional Kendall's $\tau$ at time $t$, obtained from simulations of the fitted model, is plotted against $t$ for each pair $(U_i,U_j), i,j \in \{1, \ldots, 6\},i<j$ (black line). The blue region is constructed from the $5\%$ and $95\%$ empirical quantiles of the 100 empirical estimates of Kendall's $\tau$. True Kendall's $\tau$ values determined by simulating from the true model 500 times are added in red.}
\label{fig:sim_us_runa}
\end{figure}

We use the same 100 simulated data sets as in the case with known tree structure but allow here for structure selection within our procedure. As already mentioned, evaluating our results is not straightforward in this case. Our estimated model may contain pair copulas for which we do not know the true copula families and Kendall's $\tau$ values directly. In this case we simulate 500 times from the true model and from the estimated model. Then we can calculate empirical Kendall's $\tau$ values for each of the $\frac{6\cdot 5}{2} = 15$ pairs $(U_1, U_2), (U_1, U_3), \ldots$. We compare trajectories of the empirical Kendall's $\tau$ values in Figures \ref{fig:sim_us_run1} and \ref{fig:sim_us_runa}. These trajectories look similar for the true and the estimated models. We see that also our procedure with structure selection is able to recover the simulated trajectories of Kendall's $\tau$.

For further evaluation of the proposed procedure we compare log-likelihoods of estimated and true models, as in \cite{gruber2015sequential}. 
To save computation time, we evaluate the likelihoods of estimated models based on point estimates (marginal posterior mode estimates) of the parameters, instead of evaluating the likelihoods for all posterior draws. The average log-likelihood of models without structure selection was $94\%$ of the log-likelihood of the true model, whereas the log-likelihood of the models estimated with structure selection was on average $89\%$ of the log-likelihood of the true model. It is not surprising that we perform a bit better if we assume the vine structure to be known. But the difference is not very big and in both cases, with and without vine structure selection, we obtain reasonable results. For further comparison we also estimated dynamic C-vine and D-vine copulas. 
Therefore we just restrict our structure selection procedure in Section \ref{sec:seqest} to C-vine and D-vine structures, respectively.
The dynamic C-vine and D-vine copulas achieved  $86\%$ and $88\%$ of the log-likelihood of the true model, respectively. Thus, in this scenario, allowing for general vine structures improves the fit compared to restricting the structure to C-vines or D-vines.

\section{Application: Dynamic exchange rates dependence}
\label{sec:app}
We employ the proposed dynamic vine copula model to model the dependence among 21 exchange rates with respect to the US Dollar (USD). For this we use data obtained from the FRED database of the Federal Reserve Bank of St. Louis (https://fred.stlouisfed.org/categories/94) which comprises daily log returns of 21 exchange rates with respect to the USD from 2007 to 2018, resulting in 3130 observations. The 21 currencies and their ticker symbols are summarized in Appendix \ref{sec:app_exr}.
We estimate our model based on the first $1500$ observations and evaluate its predictive performance based on the remaining 1630 observations. First the data is demeaned based on the first 1500 observations and we collect the demeaned log returns in the data matrix $Y =(y_{tj})_{t=1, .\ldots, 3130, j=1, \ldots, 21} \in \mathbb{R}^{3130 \times 21}$. 

\noindent
For the marginals we use skew Student t stochastic volatility models, i.e. we assume that
\begin{equation}
\begin{split}
Y_{tj} =& \exp(\frac{s_{tj}^{st}}{2}) \epsilon_{tj}^{st}\\
s_{tj}^{st} =& \mu_j^{st} + \phi_{j}^{st} (s_{t-1j}^{st} - \mu_{j}^{st}) + \sigma_{j}^{st} \eta_{tj}^{st}
\end{split}
\end{equation}
with $\eta_{tj} \sim N(0,1)$ independently, $ \mu_j^{st} \in \mathbb{R}, \phi_{j}^{st} \in (-1,1), \sigma_j^{st} \in (0,\infty)$ for $t=1, \ldots, 3130$. The error $\epsilon_{tj}$ follows marginally a standardized skew Student t distribution with skewness parameter $\alpha_j \in \mathbb{R}$ and degrees of freedom parameter $df_j \in (2, \infty)$ (see \cite{2019arXiv190210412K}). The corresponding density and distribution functions are denoted by $sst(\cdot|\alpha_j, df_j)$ and $SST(\cdot|\alpha_j, df_j)$, respectively. The joint distribution among the errors is modeled by the proposed dynamic vine copula model.

We follow ideas of the two step approach, commonly used in copula modeling, and assume independence among the errors $\epsilon_{tj}^{st}$ for estimating the margins. But
instead of collapsing parameters of the marginal skew Student t stochastic volatility models to point estimates and obtain the copula data based on these point estimates we follow ideas from Section \ref{sec:seqest}. For each of the 21 marginal time series $y_{1j}, \ldots, y_{1500j}$ we estimate a skew Student t stochastic volatility model as explained in \cite{2019arXiv190210412K}. The sampler is run for $1100 \cdot 25$ iterations and then we thin the samples with factor 25. The parameter draws of the skewness parameters, of the degrees of freedom parameters and of the latent log variances are denoted by $(\alpha_j^{st})^r, (df_j^{st})^r, (s_{0j}^{st})^r, \ldots, (s_{1500j}^{st})^r, r=1, \ldots, 1100$.
For each parameter draw, we obtain pseudo copula data as follows
\begin{equation}
u^r_{tj} = SST\left({y_{tj}}\exp(-\frac{(s_{tj}^{st})^r}{2})\Big|(\alpha_j^{st})^r, (df_j^{st})^r\right)
\end{equation}
for $t=1, \ldots, 1500$, $j=1, \ldots, 21$, $r=1, \ldots, 1100$.

Based on these pseudo copula data sets we fit a dynamic vine copula model. The algorithm of Section \ref{sec:seqest} is slightly modified. We start with Step 3. and set $l=1$ since we fit our model with a collection of copula data sets instead of only one copula data set. Further we set $R=1100, k=25$ and $burnin=100$. The copula families are selected from the following set $\mathcal{M} =\{ $Independence, Gaussian, Student t(df=2), Student t(df=4), Student t(df=8), eGumbel, eClayton$\}$. The estimated dynamic vine copula model is analyzed in more detail in the following.

\begin{figure}[H]
\centerline{%
\includegraphics[trim={0cm 0cm 0cm 0cm},width=0.6\textwidth]{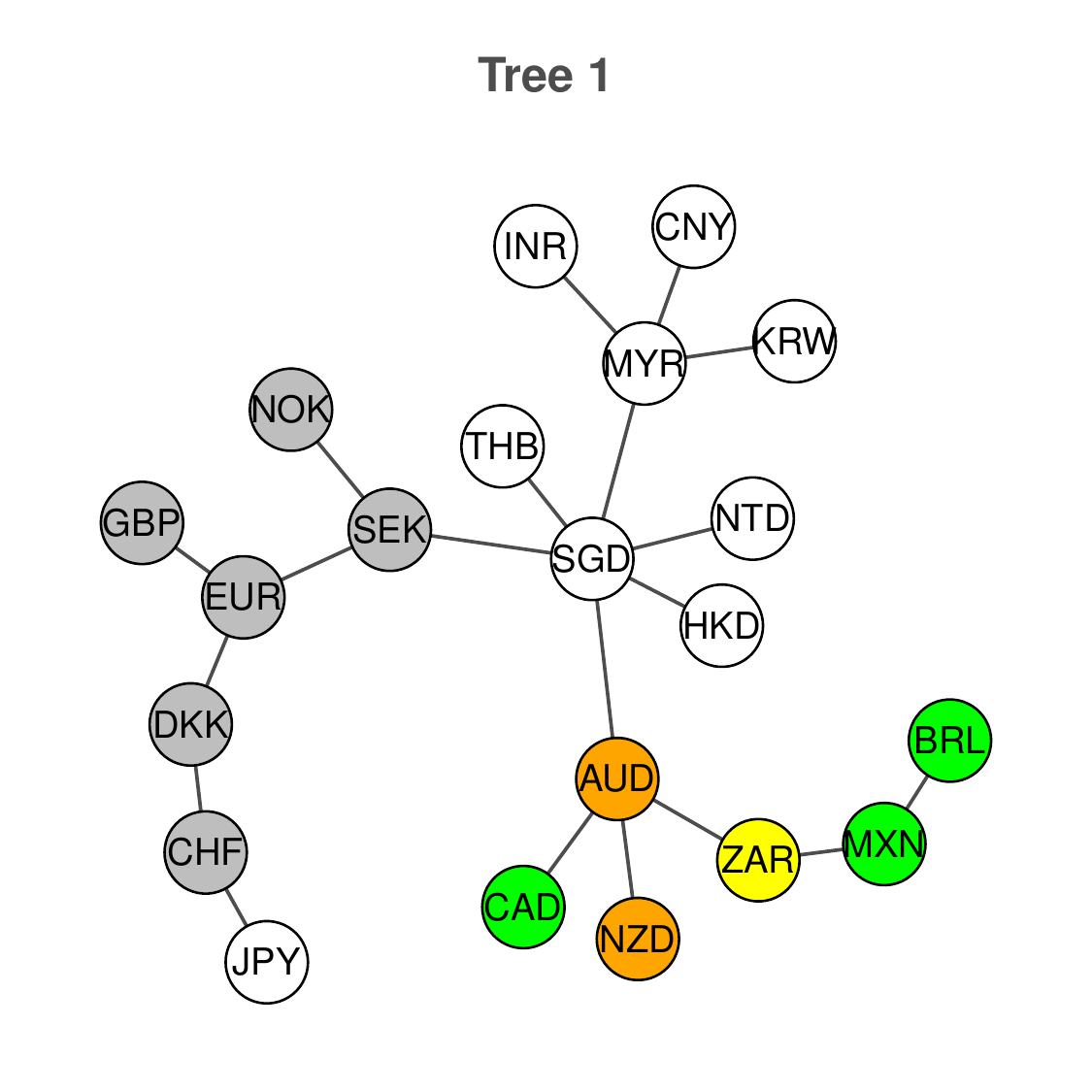}%
}%
\caption{The first tree of the vine tree structure selected for the 21-dimensional exchange rates data set. Nodes which belong to the same region have the same color (Europe: grey, Asia: white, America: green, Australasia: orange, Africa: yellow). }
\label{fig:vintree1}
\end{figure}

The first tree of the selected vine tree structure is shown in Figure \ref{fig:vintree1}.
We see that some currencies that are connected by an edge are from countries of the same region. For example, the currencies GBP/USD (British Pound to USD) and DKK/USD (Danish Krone to USD) are connected to the EUR/USD (Euro to USD). Since the vine structure is selected as the maximum spanning tree, where the absolute value of Kendall's $\tau$ serves as weight, this indicates high dependence among those currencies.
Further, we see that the selected vine structure is neither a C-vine nor a D-vine structure. The generalization of C-vine and D-vine structures to R-vine structures seems to be necessary.

In Table \ref{tab:tautypeTreelevel} we show the selected types of dependence per tree level. Above tree nine, all selected copulas are equal to the independence copula, i.e. the type of dependence is estimated to be zero. Further we see that only few static copulas were selected. The number of dynamic copulas selected decreases as we move up to higher tree levels. In total  $\frac{20\cdot 21}{2} = 210$ pair copulas are estimated, of which $150$ were set equal to the independence copula. Our proposed procedure is able to detect sparse structures. Note that the level of sparsity can be increased by adjusting the selection of the type of dependence accordingly. As mentioned in Section \ref{sec:WAIC}, we decide for the more complex type of dependence if the WAIC of the more complex model is at least  2 standard errors smaller. By increasing 2 to for example 4 standard errors, we achieve more sparsity. This might be interesting in higher dimensional settings. Since for most pair copulas in the first trees the dynamic type of dependence is selected, a vine copula model with static dependence might not be appropriate for those selected pair copula terms with time-varying dependence.

\begin{table}[ht]
\centering
\begin{tabular}{r|rrr}
  \hline
Tree & Dynamic & Static & Zero \\ 
  \hline
1 &  18 &   2 &   0 \\ 
  2 &  16 &   1 &   2 \\ 
  3 &   7 &   0 &  11 \\ 
  4 &   6 &   1 &  10 \\ 
  5 &   2 &   0 &  14 \\ 
  6 &   3 &   0 &  12 \\ 
  7 &   2 &   0 &  12 \\ 
  8 &   1 &   0 &  12 \\ 
  9 &   1 &   0 &  11 \\ 
   \hline
\end{tabular}
\caption{We show how often the different types of dependence (dynamic, static, zero) were selected per tree level for the first nine trees.}
\label{tab:tautypeTreelevel}
\end{table}

\begin{figure}[H]
\centerline{%
\includegraphics[trim={0cm 6.25cm 0cm 0cm},width=0.9\textwidth]{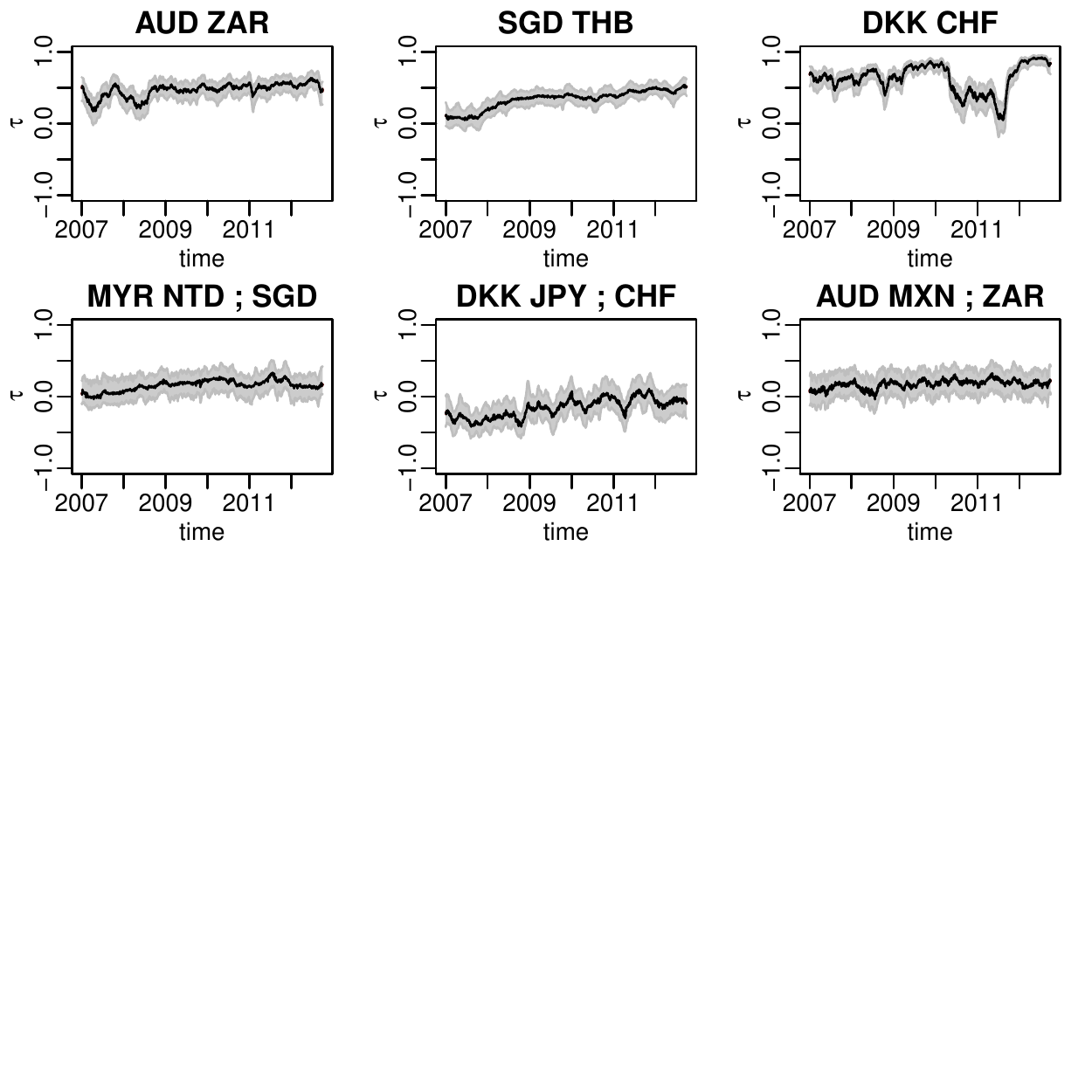}%
}%
\caption{Visualization of the dynamic Kendall's $\tau$ for some chosen pair copulas of the dynamic vine copula model estimated for the 21-dimensional exchange rates data set. Rows 1 and 2 of this plot correspond to pair copulas in Trees 1 and 2, respectively. The black line shows marginal posterior mode estimates of Kendall's $\tau$ plotted against time $t$. A $90\%$ credible region constructed from the estimated $5\%$ and $95\%$ posterior quantiles is added in grey.}
\label{fig:dyntau_tree1to3}
\end{figure}

Figure \ref{fig:dyntau_tree1to3} shows how the dynamic Kendall's $\tau$ values evolve over time.
We see that the dependence between the exchange rates AUD/USD (Australian Dollar to USD) and ZAR/USD (South African Rand to USD) varies more in 2007 and 2008, during the financial crisis, and remains almost constant after that period. Further we observe that the Kendall's $\tau$ between SGD/USD (Singapore Dollar to USD) and THB/USD (Thai Baht to USD) is close to zero in 2007 and then starts to increase after 2007. The dependence between DKK/USD (Danish Krone to USD) and CHF/USD (Swiss Franc to USD) is rather high but decreases in 2010 and reaches its lowest point in 2011.
This might be the effect of the introduction of the cap on the Swiss Franc on 6 September 2011 by the Swiss National Bank. The minimum exchange rate was set at 1.2 CHF (Swiss Franc) per EUR (Euro).
The second row of Figure \ref{fig:dyntau_tree1to3} shows fitted conditional Kendall's $\tau$ values. For example, we see how the Kendall'$\tau$ of the exchange rates DKK/USD (Danish Krone to USD) and JPY/USD (Japanese Yen to USD) conditional on CHF/USD (Swiss Franc to USD) evolves over time. The conditional dependence varies between $-0.5$ and $0$. 
We also provide quantification of uncertainty of the Kendall's $\tau$ values through credible intervals. This is an advantage of our Bayesian approach compared to the frequentist approach of \cite{almeida2016modeling} for dynamic D-vine copulas, where uncertainty quantification was not provided.

As already mentioned, the proposed dynamic vine copula model can be seen as a generalization of static vine copulas and as a generalization of the dynamic C-vine and D-vine copula models of \cite{goel2019analyzing} and \cite{almeida2016modeling}.
We would like to further support the hypothesis that our proposed dynamic vine copula model is a needed generalization and is able to describe the dependence structure more appropriate than its competitor models: a dynamic C-vine,  a dynamic D-vine and a static vine copula. Therefore we compare these models with respect to their predictive accuracy.
To estimate the dynamic C-vine and D-vine copula model, we adjust our structure selection procedure in Section \ref{sec:seqest} accordingly. The families are selected from the same set $\mathcal{M}$ that we used for the dynamic vine copula. The static vine copula model is estimated with the algorithm of \cite{dissmann2013selecting}, as implemented in the R-package \texttt{rvinecopulib} (\cite{nagler2018rvinecopulib}). Here we allow for all parametric copula families that are implemented in the \texttt{rvinecopulib} package.  For all models we use skew Student t stochastic volatility  models for the margins. For the static vine copula model the pseudo copula data is obtained by fixing the parameters of the skew Student t stochastic volatility models at marginal posterior mode estimates. The competitor models are also estimated based on the first 1500 observations of our data. For all models we obtain one day ahead predictive scores for the other 1630 days in our data set. We proceed as in \cite{2019arXiv190210412K}. Instead of refitting the models 1629 times, we keep static model parameters fixed at point estimates (marginal posterior mode estimates for the dynamic models, maximum likelihood estimates for the static vine copula) and only update dynamic parameters. This reduces computation time a lot. To further reduce computation time, we evaluate the log predictive likelihoods at point estimates instead of averaging over all posterior draws. Similar to  \cite{kastner2019sparse},  we call this pseudo log predictive scores (plps).
The plps at time $t>1500$ has the following structure
\begin{equation}
\text{plps}_t(y_1, \ldots, y_{21}) = \ln(\hat c_t({\hat u_1, \ldots, \hat u_{21}})) + \sum_{j=1}^{21} \ln\left(sst\left({y_{j}}\exp(-\frac{\hat s_{tj}^{st}}{2})\Big|\hat\alpha_j^{st}, \hat{df}_j^{st}\right)\right)-\frac{\hat s_{tj}^{st}}{2},
\end{equation}
where $\hat c_t$ is the estimated copula density of one of the four considered models obtained by fixing the corresponding parameters at point estimates (marginal posterior mode estimates for the dynamic models, maximum likelihood estimates for the static vine copula). Further $\hat u_j=$ \newline $SST\left({y_{j}}\exp(-\frac{\hat s_{tj}^{st}}{2})\Big|\hat\alpha_j^{st}, \hat{df}_j^{st}\right)$ with marginal posterior mode estimates $\hat s_{tj}^{st}, \hat\alpha_j^{st}, \hat{df}_j^{st} $ for $j=1, \ldots, 21$. The marginal contribution $\sum_{j=1}^{21} \ln\left(sst\left({y_{j}}\exp(-\frac{\hat s_{tj}^{st}}{2})\Big|\hat\alpha_j^{st}, \hat{df}_j^{st}\right)\right)-\frac{\hat s_{tj}^{st}}{2}$ is the same for all considered models. So we compare the models with respect to the copula contributions $\ln(\hat c({\hat u_1, \ldots, \hat u_{21}}))$ to which we refer as copula plps. Note that a higher (copula) plps is an indication for better forecasting accuracy.

 Table \ref{tab:copLP} shows the cumulative copula plps, i.e. the sum over all 1630 plps. We see that the two vine copula models with flexible tree structure outperform the dynamic C-vine and D-vine copulas.  Further the dynamic vine copula, for which the selected structure deviates clearly from a C-vine and a D-vine structure and for which many pair copulas have a dynamic type of dependence provides the most accurate forecasts. Our conclusion is that the dynamic vine copula model provides a useful generalization of static vine copula models as well as of dynamic C-vine and D-vine copula models.


\begin{table}[ht]
\centering
\begin{tabular}{rrrrr}
  \hline
 & Dynamic vine &Dynamic C-vine &  Dynamic D-vine & Static vine \\ 
  \hline
     
copula plps & \textbf{11643} & 11132  & 11126 &  11267\\ 
   \hline
\end{tabular}
\caption{Cumulative one day ahead copula plps for the four considered models: Dynamic vine, dynamic C-vine, dynamic D-vine and static vine copula.}
\label{tab:copLP}
\end{table}

\section{Conclusion and future research}
\label{sec:conc}
We introduced a class of dynamic vine copula models and provided a novel Bayesian estimation procedure based on an approximation of the posterior distribution allowing for simplification of the sampler.
Here we allowed for the selection of the pair copula family, the selection of the type of dependence for each pair copula term and the sequential selection of a static (time-constant) vine structure.
 The application showed that the dynamic vine copula model is  a useful extension of static vine copulas and of dynamic C-vine and D-vine copulas.

Our estimation procedure propagates uncertainty of copula parameter estimation from lower to higher trees. But the type of dependence is selected with WAIC and then fixed before we move to the next tree. Instead of using the WAIC or any other information criteria, shrinkage priors as proposed by \cite{bitto2019achieving} that allow to shrink dynamic parameters to static ones might be an interesting alternative to be studied in future research.

One restriction of our approach is that both the vine tree structure as well as the pair copula families are assumed not to change over time. In the future we are interested in overcoming these restrictions. 

Further, it would be interesting to study in more detail how the dynamic vine copula model performs in situations, where appropriate dependence modeling is crucial, for example in financial risk management. The dynamic vine copula model might lead to more accurate value at risk predictions than those obtained from static vine copula models. 
Another example is pairs trading.  \cite{stubinger2018statistical} showed that profitable trading strategies can be constructed with static vine copula models. These strategies might be improved by allowing for dynamic dependencies.

Lastly, we think that the ability of the vine copula framework to scale bivariate copula models to copula models of arbitrary dimensions has not been fully exploited yet. We are sure that there is a variety of useful extensions of static vine copula models that build on sophisticated bivariate copula models. For example, one could allow for bivariate copula families with more than one parameter such as the BB1 family or for bivariate dynamic mixture copulas as studied by \cite{2019arXiv190210412K}, where both mixture components share the same dynamic on Kendall's $\tau$. Alternatively one could also study bivariate mixture copula models with one dynamic and one static component.  

\appendix

\section{Additional material for parameter sharing (Section \ref{sec:bivdynmod})}
\label{sec:app_ps}

\begin{figure}[H]
\centerline{%
\includegraphics[trim={0cm 0cm 0cm 0cm},width=0.6\textwidth]{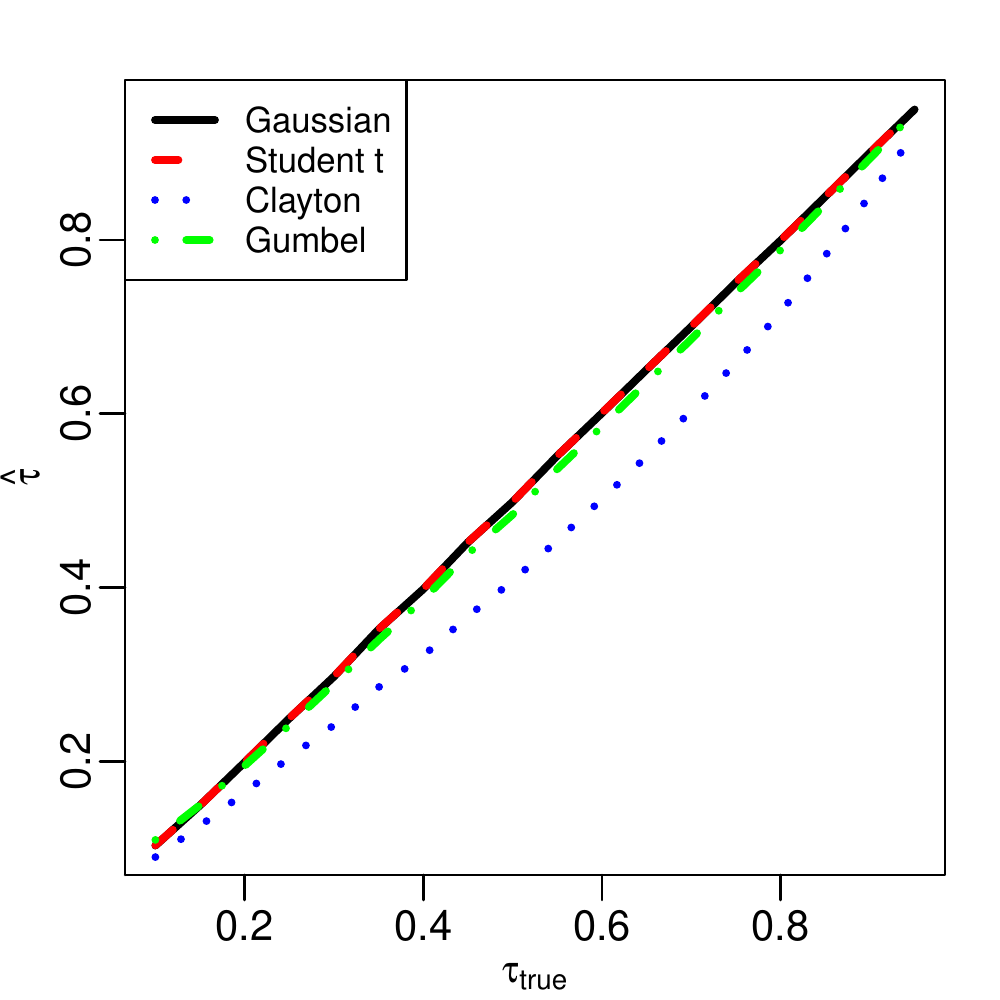}%
}%
\caption{This plot shows average Kendall's $\tau$ estimates, $\hat\tau$, for different copula families (Gaussian, Student t(df=4), Clayton, Gumbel) plotted against the Kendall's $\tau$ that was used for simulation, $\tau_{true}$.}
\label{fig:sharetau}
\end{figure}

Our procedure in Section \ref{sec:bivdynmod} shares parameters among different copula families. This is motivated by the fact that the parameter $s_t$, the Fisher's Z transform of Kendall's $\tau$, is similar for different copula families. To support this statement, we conduct the following experiment: We simulate 100 bivariate data sets, each containing 1000 observations, from the bivariate Student t copula with 4 degrees of freedom and copula parameter $\rho_{true}$. The corresponding Kendall's $\tau$ is obtained as $\tau_{true}=\frac{2}{\pi}\arcsin(\rho_{true})$. For each data set, we estimate the copula parameter of the Gaussian, Student t, Clayton and Gumbel copula by maximizing the likelihood and transform the estimates to the corresponding Kendall's $\tau$ values. We obtain 100 estimated Kendall's $\tau$ values for each copula family and take the average of those 100 values, which we denote by $\hat \tau$. This results in four different $\hat \tau$ values corresponding to four different copula families. This procedure is repeated for different values of $\tau_{true}$ and the average Kendall's $\tau$ estimate, $\hat\tau$, is shown in Figure \ref{fig:sharetau} for each value of $\tau_{true}$.  We see that the estimated Kendall's $\tau$ values for the Gaussian, Student t and Gumbel copula are very close to each other. Although the Kendall's $\tau$ estimates for the Clayton copula are a bit further apart, we think that they are still reasonable close such that parameter sharing is justified.

\section{Parameter specification for the simulation study in Section 3.3}
\label{sec:app_sim}
Parameters of a dynamic vine copula model are here specified through matrices. The last row shows parameters corresponding to pair copulas in the first tree, the second last row parameters corresponding to pair copulas in the second tree and so on. If we set the dispersion parameter and the standard deviation parameter to zero, we obtain a static copula model. 
$$
\mu =
\begin{pmatrix}{}
   &  &  &  &  &  \\ 
  0.0 &  &  &  &  &  \\ 
  0.0 & 0.0 &  &  &  &  \\ 
  0.0 & 0.3 & 0.0 &  &  &  \\ 
  0.3 & 0.4 & 0.3 & 0.0 &  &  \\ 
  0.9 & 0.6 & 0.8 & 0.8 & 1.0 &  \\ 
  \end{pmatrix}
  \phi = 
\begin{pmatrix}{}
   &  &  &  &  &  \\ 
  0.00 &  &  &  &  &  \\ 
  0.00 & 0.00 &  &  &  &  \\ 
  0.00 & 0.00 & 0.00 &  &  &  \\ 
  0.98 & 0.90 & 0.00 & 0.00 &  &  \\ 
  0.95 & 0.98 & 0.90 & 0.00 & 0.00 &  \\ 
  \end{pmatrix}
$$

$$
\sigma = 
\begin{pmatrix}{}
   &  &  &  &  &  \\ 
  0.00 &  &  &  &  &  \\ 
  0.00 & 0.00 &  &  &  &  \\ 
  0.00 & 0.00 & 0.00 &  &  &  \\ 
  0.05 & 0.10 & 0.00 & 0.00 &  &  \\ 
  0.10 & 0.03 & 0.05 & 0.00 & 0.00 &  \\ 
  \end{pmatrix}
  $$
  
  $$
  \text{family} = 
  \begin{pmatrix}{}
   &  &  &  &  &  \\ 
  \text{Independence} &  &  &  &  &  \\ 
  \text{Independence} & \text{Independence} &  &  &  &  \\ 
  \text{Independence} & \text{eClayton} & \text{Independence} &  &  &  \\ 
  \text{Gaussian} & \text{Student t(df=4)} & \text{eGumbel} & \text{Independence} &  &  \\ 
  \text{Gaussian} & \text{Student t(df=4)} & \text{eClayton} & \text{eGumbel} & \text{Gaussian} &  \\ 
  \end{pmatrix}
  $$
  
\noindent 
Note that, within the dynamic bivariate copula model, the stationary distribution of the AR(1) process is given by
\begin{equation*}
s|\mu, \phi, \sigma \sim N\left(\mu, \frac{\sigma^2}{1-\phi^2}\right).
\end{equation*}  
for  a state $s$.
Using the density transformation rule this implies the following density for Kendall's $\tau$ (the Fisher's Z transform of $s$)
\begin{equation}
f(\tau|\mu, \phi, \sigma) = \varphi\left(F_Z(\tau)|\mu, \frac{\sigma^2}{1-\phi^2}\right) \frac{1}{1-\tau^2}, \tau \in (-1,1).
\label{eq:taudens}
\end{equation}

\begin{figure}[H]
\centerline{%
\includegraphics[trim={0cm 0cm 0cm 0cm},width=0.6\textwidth]{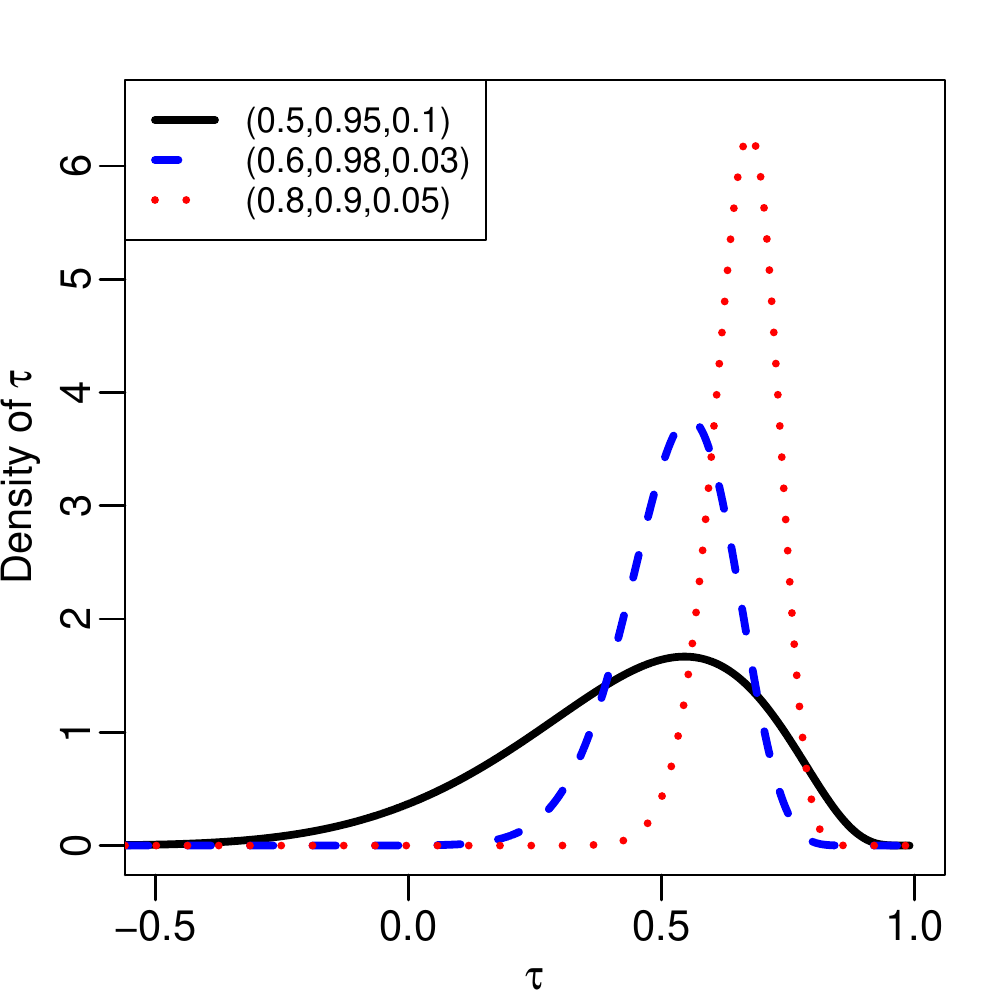}%
}%
\caption{We show the stationary density of $\tau$ given in \eqref{eq:taudens} for different values of $\mu, \phi$ and $\sigma$. Each line is associated with a vector ($\mu, \phi, \sigma$) given in the legend.}
\label{fig:taudens}
\end{figure}

To obtain an understanding of what different choices of $\mu$, $\phi$ and $\sigma$ imply for $\tau$, we show the density given in \eqref{eq:taudens} in Figure \ref{fig:taudens}. We consider the values of $\mu$, $\phi$ and $\sigma$ which are used in the first tree.

 \newpage
\section{Exchange rates (to the US Dollar) data set}
  \label{sec:app_exr}
\begin{table}[H]
\centering
\begin{tabular}{l|l}
\hline
Ticker & Currency \\
\hline
BRL & Brazilian Real \\ 
CAD & Canadian Dollar \\ 
CNY & Chinese Yuan \\ 
DKK & Danish Krone \\ 
HKD & Hong Kong Dollar \\ 
INR & Indian Rupees \\ 
JPY & Japanese Yen \\ 
KRW & South Korean Won \\ 
MYR & Malaysian Ringgit \\ 
MXN & Mexican New Pesos \\ 
NOK & Norwegian Krone \\ 
SEK & Swedish Krona \\ 
ZAR & South African Rand \\ 
SGD & Singapore Dollar \\ 
CHF & Swiss Franc \\ 
NTD & New Taiwan Dollar \\ 
THB & Thai Baht \\ 
AUD & Australian Dollar \\ 
EUR & Euro \\ 
NZD & New Zealand Dollar \\ 
GBP &  British Pound \\
\hline
\end{tabular}
\caption{The 21 currencies with corresponding ticker symbols used in the application in Section 4.}
\label{tab:copLP}
\end{table}  

\vspace*{0.25cm}
\subsubsection*{Acknowledgements}
The second author is supported by  the  German  Research  Foundation  (DFG grant
CZ 86/4-1). Computations were performed on a Linux cluster supported by DFG grant INST 95/919-1 FUGG.

\bibliographystyle{spbasic}    
\bibliography{References}{}

\begin{thebibliography}{49}
\providecommand{\natexlab}[1]{#1}
\providecommand{\url}[1]{{#1}}
\providecommand{\urlprefix}{URL }
\expandafter\ifx\csname urlstyle\endcsname\relax
  \providecommand{\doi}[1]{DOI~\discretionary{}{}{}#1}\else
  \providecommand{\doi}{DOI~\discretionary{}{}{}\begingroup
  \urlstyle{rm}\Url}\fi
\providecommand{\eprint}[2][]{\url{#2}}

\bibitem[{Aas(2016)}]{aas2016pair}
Aas K (2016) Pair-copula constructions for financial applications: A review.
  Econometrics 4(4):43

\bibitem[{Aas et~al(2009)Aas, Czado, Frigessi, and Bakken}]{aas2009pair}
Aas K, Czado C, Frigessi A, Bakken H (2009) Pair-copula constructions of
  multiple dependence. Insurance: Mathematics and Economics 44(2):182--198

\bibitem[{Acar et~al(2019)Acar, Czado, and Lysy}]{acar2019flexible}
Acar EF, Czado C, Lysy M (2019) Flexible dynamic vine copula models for
  multivariate time series data. Econometrics and Statistics 12:181--197

\bibitem[{Almeida and Czado(2012)}]{almeida2012efficient}
Almeida C, Czado C (2012) Efficient Bayesian inference for stochastic
  time-varying copula models. Computational Statistics \& Data Analysis
  56(6):1511--1527

\bibitem[{Almeida et~al(2016)Almeida, Czado, and Manner}]{almeida2016modeling}
Almeida C, Czado C, Manner H (2016) Modeling high-dimensional time-varying
  dependence using dynamic D-vine models. Applied Stochastic Models in Business
  and Industry 32(5):621--638

\bibitem[{Baele et~al(2010)Baele, Bekaert, and
  Inghelbrecht}]{baele2010determinants}
Baele L, Bekaert G, Inghelbrecht K (2010) The determinants of stock and bond
  return comovements. The Review of Financial Studies 23(6):2374--2428

\bibitem[{Barthel et~al(2018)Barthel, Geerdens, Czado, and
  Janssen}]{barthel2018dependence}
Barthel N, Geerdens C, Czado C, Janssen P (2018) {Dependence modeling for
  recurrent event times subject to right-censoring with D-vine copulas}.
  Biometrics 75:439--451

\bibitem[{Bedford and Cooke(2001)}]{bedford2001probability}
Bedford T, Cooke RM (2001) Probability density decomposition for conditionally
  dependent random variables modeled by vines. Annals of Mathematics and
  Artificial Intelligence 32(1-4):245--268

\bibitem[{Bitto and Fr{\"u}hwirth-Schnatter(2019)}]{bitto2019achieving}
Bitto A, Fr{\"u}hwirth-Schnatter S (2019) Achieving shrinkage in a time-varying
  parameter model framework. Journal of Econometrics 210(1):75--97

\bibitem[{Brechmann and Czado(2013)}]{brechmann2013risk}
Brechmann EC, Czado C (2013) Risk management with high-dimensional vine
  copulas: An analysis of the Euro Stoxx 50. Statistics \& Risk Modeling
  30(4):307--342

\bibitem[{Brechmann et~al(2012)Brechmann, Czado, and
  Aas}]{brechmann2012truncated}
Brechmann EC, Czado C, Aas K (2012) Truncated regular vines in high dimensions
  with application to financial data. Canadian Journal of Statistics
  40(1):68--85

\bibitem[{Czado(2019)}]{czadoanalyzing}
Czado C (2019) Analyzing Dependent Data with Vine Copulas. Lecture Notes in
  Statistics, Springer

\bibitem[{Dissmann et~al(2013)Dissmann, Brechmann, Czado, and
  Kurowicka}]{dissmann2013selecting}
Dissmann J, Brechmann EC, Czado C, Kurowicka D (2013) Selecting and estimating
  regular vine copulae and application to financial returns. Computational
  Statistics \& Data Analysis 59:52--69

\bibitem[{Embrechts et~al(2002)Embrechts, McNeil, and
  Straumann}]{embrechts2002correlation}
Embrechts P, McNeil A, Straumann D (2002) Correlation and dependence in risk
  management: properties and pitfalls. Risk Management: Value at Risk and
  Beyond 1:176--223

\bibitem[{Engle(2002)}]{engle2002dynamic}
Engle R (2002) Dynamic conditional correlation: A simple class of multivariate
  generalized autoregressive conditional heteroskedasticity models. Journal of
  Business \& Economic Statistics 20(3):339--350

\bibitem[{Erhardt et~al(2015)Erhardt, Czado, and Schepsmeier}]{erhardt2015r}
Erhardt TM, Czado C, Schepsmeier U (2015) R-vine models for spatial time series
  with an application to daily mean temperature. Biometrics 71(2):323--332

\bibitem[{Garthwaite et~al(2016)Garthwaite, Fan, and
  Sisson}]{garthwaite2016adaptive}
Garthwaite PH, Fan Y, Sisson SA (2016) Adaptive optimal scaling of
  Metropolis--Hastings algorithms using the Robbins--Monro process.
  Communications in Statistics-Theory and Methods 45(17):5098--5111

\bibitem[{Goel and Mehra(2019)}]{goel2019analyzing}
Goel A, Mehra A (2019) Analyzing Contagion Effect in Markets During Financial
  Crisis Using Stochastic Autoregressive Canonical Vine Model. Computational
  Economics 53(3):921--950

\bibitem[{Green(1995)}]{green1995reversible}
Green PJ (1995) Reversible jump Markov chain Monte Carlo computation and
  Bayesian model determination. Biometrika 82(4):711--732

\bibitem[{Gruber and Czado(2015)}]{gruber2015sequential}
Gruber LF, Czado C (2015) Sequential Bayesian model selection of regular vine
  copulas. Bayesian Analysis 10(4):937--963

\bibitem[{Gruber and Czado(2018)}]{gruber2018bayesian}
Gruber LF, Czado C (2018) Bayesian model selection of regular vine copulas.
  Bayesian Analysis 13(4):1107--1131

\bibitem[{Haff et~al(2010)Haff, Aas, and Frigessi}]{haff2010simplified}
Haff IH, Aas K, Frigessi A (2010) On the simplified pair-copula
  construction—simply useful or too simplistic? Journal of Multivariate
  Analysis 101(5):1296--1310

\bibitem[{Hafner and Manner(2012)}]{hafner2012dynamic}
Hafner CM, Manner H (2012) Dynamic stochastic copula models: Estimation,
  inference and applications. Journal of Applied Econometrics 27(2):269--295

\bibitem[{Harvey et~al(1994)Harvey, Ruiz, and
  Shephard}]{harvey1994multivariate}
Harvey A, Ruiz E, Shephard N (1994) Multivariate stochastic variance models.
  The Review of Economic Studies 61(2):247--264

\bibitem[{Joe(2014)}]{joe2014dependence}
Joe H (2014) Dependence modeling with copulas. CRC Press

\bibitem[{Kastner(2016)}]{kastner2016dealing}
Kastner G (2016) Dealing with stochastic volatility in time series using the R
  package stochvol. Journal of Statistical Software 69(5):1--30

\bibitem[{Kastner(2019)}]{kastner2019sparse}
Kastner G (2019) Sparse Bayesian time-varying covariance estimation in many
  dimensions. Journal of Econometrics 210(1):98--115

\bibitem[{Kastner et~al(2017)Kastner, Fr{\"u}hwirth-Schnatter, and
  Lopes}]{kastner2017efficient}
Kastner G, Fr{\"u}hwirth-Schnatter S, Lopes HF (2017) Efficient Bayesian
  inference for multivariate factor stochastic volatility models. Journal of
  Computational and Graphical Statistics 26(4):905--917

\bibitem[{{Kreuzer} and {Czado}(2019)}]{2019arXiv190210412K}
{Kreuzer} A, {Czado} C (2019) {Efficient Bayesian inference for univariate and
  multivariate non linear state space models with univariate autoregressive
  state equation}. arXiv e-prints arXiv:1902.10412

\bibitem[{Liu(1994)}]{liu1994collapsed}
Liu JS (1994) The collapsed Gibbs sampler in Bayesian computations with
  applications to a gene regulation problem. Journal of the American
  Statistical Association 89(427):958--966

\bibitem[{Min and Czado(2011)}]{min2011bayesian}
Min A, Czado C (2011) Bayesian model selection for D-vine pair-copula
  constructions. Canadian Journal of Statistics 39(2):239--258

\bibitem[{M{\"o}ller et~al(2018)M{\"o}ller, Spazzini, Kraus, Nagler, and
  Czado}]{moller2018vine}
M{\"o}ller A, Spazzini L, Kraus D, Nagler T, Czado C (2018) Vine copula based
  post-processing of ensemble forecasts for temperature. arXiv preprint
  arXiv:181102255

\bibitem[{Morales-N{\'a}poles(2010)}]{morales2010counting}
Morales-N{\'a}poles O (2010) Counting vines. In: Dependence modeling: Vine
  copula handbook, World Scientific, pp 189--218

\bibitem[{Murray et~al(2010)Murray, Adams, and MacKay}]{murray2010elliptical}
Murray I, Adams RP, MacKay DJ (2010) Elliptical slice sampling. Proceedings of
  the 13th International Conference on Artificial Intelligence and Statistics
  (AISTATS) 9:541--548

\bibitem[{Nagler and Vatter(2018)}]{nagler2018rvinecopulib}
Nagler T, Vatter T (2018) rvinecopulib: High performance algorithms for vine
  copula modeling. R package version 02 8(0)

\bibitem[{Oh and Patton(2018)}]{oh2018time}
Oh DH, Patton AJ (2018) Time-varying systemic risk: Evidence from a dynamic
  copula model of cds spreads. Journal of Business \& Economic Statistics
  36(2):181--195

\bibitem[{Pitt and Shephard(1999)}]{pitt1999time}
Pitt M, Shephard N (1999) Time varying covariances: a factor stochastic
  volatility approach. Bayesian Statistics 6:547--570

\bibitem[{Richard and Zhang(2007)}]{richard2007efficient}
Richard JF, Zhang W (2007) Efficient high-dimensional importance sampling.
  Journal of Econometrics 141(2):1385--1411

\bibitem[{Sklar(1959)}]{sklar1959fonctions}
Sklar M (1959) Fonctions de repartition an dimensions et leurs marges. Publ
  inst statist univ Paris 8:229--231

\bibitem[{Stoeber et~al(2013)Stoeber, Joe, and Czado}]{stoeber2013simplified}
Stoeber J, Joe H, Czado C (2013) Simplified pair copula
  constructions—limitations and extensions. Journal of Multivariate Analysis
  119:101--118

\bibitem[{St{\"u}binger et~al(2018)St{\"u}binger, Mangold, and
  Krauss}]{stubinger2018statistical}
St{\"u}binger J, Mangold B, Krauss C (2018) Statistical arbitrage with vine
  copulas. Quantitative Finance 18(11):1831--1849

\bibitem[{Tan et~al(2019)Tan, Panagiotelis, and
  Athanasopoulos}]{tan2019bayesian}
Tan BK, Panagiotelis A, Athanasopoulos G (2019) Bayesian inference for the
  one-factor copula model. Journal of Computational and Graphical Statistics
  28(1):155--173

\bibitem[{Van~Dyk and Jiao(2015)}]{van2015metropolis}
Van~Dyk DA, Jiao X (2015) Metropolis-Hastings within partially collapsed Gibbs
  samplers. Journal of Computational and Graphical Statistics 24(2):301--327

\bibitem[{Vatter and Chavez-Demoulin(2015)}]{vatter2015generalized}
Vatter T, Chavez-Demoulin V (2015) Generalized additive models for conditional
  dependence structures. Journal of Multivariate Analysis 141:147--167

\bibitem[{Vatter and Nagler(2018)}]{vatter2018generalized}
Vatter T, Nagler T (2018) Generalized additive models for pair-copula
  constructions. Journal of Computational and Graphical Statistics
  27(4):715--727

\bibitem[{Vehtari et~al(2017)Vehtari, Gelman, and Gabry}]{vehtari2017practical}
Vehtari A, Gelman A, Gabry J (2017) Practical Bayesian model evaluation using
  leave-one-out cross-validation and WAIC. Statistics and Computing
  27(5):1413--1432

\bibitem[{Wainwright et~al(2008)Wainwright, Jordan
  et~al}]{wainwright2008graphical}
Wainwright MJ, Jordan MI, et~al (2008) Graphical models, exponential families,
  and variational inference. Foundations and Trends{\textregistered} in Machine
  Learning 1(1--2):1--305

\bibitem[{Watanabe(2010)}]{watanabe2010asymptotic}
Watanabe S (2010) Asymptotic equivalence of Bayes cross validation and widely
  applicable information criterion in singular learning theory. Journal of
  Machine Learning Research 11(Dec):3571--3594

\bibitem[{Yu and Meng(2011)}]{yu2011center}
Yu Y, Meng XL (2011) To center or not to center: That is not the question—an
  Ancillarity--Sufficiency Interweaving Strategy (ASIS) for boosting MCMC
  efficiency. Journal of Computational and Graphical Statistics 20(3):531--570

\end{thebibliography}

\end{document}